\documentclass[final,twocolumn]{IEEEtran}
\usepackage[latin9]{inputenc}
\usepackage{color}
\usepackage{float}
\usepackage{amsmath}
\usepackage{amsthm}
\usepackage{amssymb}
\usepackage{graphicx}
\usepackage{esint}

\makeatletter
\theoremstyle{plain}
\newtheorem{thm}{\protect\theoremname}
\theoremstyle{plain}
\newtheorem{prop}[thm]{\protect\propositionname}

\usepackage{eurosym}
\usepackage{cite}
\usepackage{algorithmic}\usepackage{array}\usepackage{url}\usepackage{graphicx}
\usepackage[percent]{overpic}
\usepackage{amsfonts}
\usepackage{times}
\usepackage{amsthm}

\ifCLASSOPTIONcompsoc
\else
\fi
\ifCLASSOPTIONcaptionsoff
\let\MYoriglatexcaption\caption
 \renewcommand{\caption}[2][\relax]{\MYoriglatexcaption[#2]{#2}}
\fi

\allowdisplaybreaks

\makeatother

\providecommand{\propositionname}{Proposition}
\providecommand{\theoremname}{Theorem}

\begin{document}

\title{Void Probabilities and Cauchy-Schwarz Divergence for Generalized
Labeled Multi-Bernoulli Models}

\author{Michael Beard, Ba-Tuong Vo, Ba-Ngu Vo, and Sanjeev Arulampalam\thanks{M. Beard is with the Defence Science and Technology Organisation,
Rockingham, WA, Australia, and also with the Department of Electrical
and Computer Engineering, Curtin University, Bentley, WA, Australia
(e-mail: michael.beard@dsto.defence.gov.au)}\thanks{B.-T. Vo and B.-N. Vo are with the Department of Electrical and Computer
Engineering, Curtin University, Bentley, WA, Australia (e-mail: ba-tuong.vo@curtin.edu.au;
ba-ngu.vo@curtin.edu.au).}\thanks{S. Arulampalam is with the Defence Science and Technology Organisation,
Edinburgh, SA, Australia (e-mail: sanjeev.arulampalam@dsto.defence.gov.au).}}
\maketitle
\begin{abstract}
The generalized labeled multi-Bernoulli (GLMB) is a family of tractable
models that alleviates the limitations of the Poisson family in dynamic
Bayesian inference of point processes. In this paper, we derive closed
form expressions for the void probability functional and the Cauchy-Schwarz
divergence for GLMBs. The proposed analytic void probability functional
is a necessary and sufficient statistic that uniquely characterizes
a GLMB, while the proposed analytic Cauchy-Schwarz divergence provides
a tractable measure of similarity between GLMBs. We demonstrate the
use of both results on a partially observed Markov decision process
for GLMBs, with Cauchy-Schwarz divergence based reward, and void probability
constraint.\end{abstract}


\section{Introduction}

Point patterns are ubiquitous in nature, for example the states of
objects in multi-object systems such as the coordinates of molecules
in a liquid/crystal, trees in a forest, stars in a galaxy and so on
\cite{Daley1988,Stoyan1995,Moller2004}. Point processes (specifically
simple finite point processes or random finite sets) are probabilistic
models for point patterns, derived from stochastic geometry - the
study of random geometrical objects ranging from collections of points
to arbitrary closed sets \cite{Stoyan1995,Molchanov2006}. Point process
theory provides the tools for characterizing the underlying laws of
the point patterns and entails a diverse range of applications areas,
such as forestry \cite{Stoyan2000}, geology \cite{Ogata1999}, biology
\cite{Marmarelis2005,Ji2009}, physics \cite{Snyder1981}, computer
vision \cite{Baddeley1993,Hoseinnezhad2012}, wireless networks \cite{Baccelli1997,Haenggi2005,Haenggi2009},
communications \cite{Biglieri2007,Angelosante2009}, multi-target
tracking \cite{Mahler2003,Mahler2014}, and robotics \cite{Mullane2011,Lee2014,Deusch2015}.

In addition to the probability distribution, the \emph{void-probability
functional} (or simply \emph{void probabilities}) is another fundamental
descriptor of a point process \cite{Daley1988,Stoyan1995,Moller2004}.
The void probability on a given region is the probability that it
contains no points of the point process. Rényi's celebrated theorem
states that the probability law of a simple point process is uniquely
determined by the void probabilities on the bounded Borel sets \cite{Daley1988,Stoyan1995,Moller2004}.
Analytic expressions for the void probabilities are available for
point processes such as Poisson and independent and identically distributed
(IID) cluster. In general, the void probabilities constitute an intuitive
and powerful descriptor that also characterizes the more general random
closed sets via Choquet's capacity theorem \cite{Matheron1975,Molchanov2006}.

Apart from characterizing point processes, measuring their similarlies/differences
is essential in the study of point patterns. Information-based divergences
are fundamental in the statistical analysis of random variables \cite{Cover1991},
and divergences such as Kullback-Leibler, Rényi, Csiszár-Morimoto
(or Ali-Silvey), and Cauchy-Schwarz have been developed for point
processes \cite{Mahler2014,Mahler1998,Mahler2004,Ristic2010,Hoang2015,Ristic2011}.
However, in general these divergences cannot be computed analytically.
Arguably the most tractable of these is the Cauchy-Schwarz divergence,
which admits a closed form expression for Poisson point processes.
Indeed the Cauchy-Schwarz divergence between two Poisson point processes
is given by half the squared $L_{2}$-distance between their intensity
functions \cite{Hoang2015}. Moreover, this result has also been extended
to mixtures of Poisson point processes \cite{Hoang2015}.

While the Poisson model enjoys many elegant analytical properties,
it is too simplistic for problems such as dynamic Bayesian inference
of point processes, and a more suitable model is the generalized labeled
multi-Bernoulli (GLMB) \cite{Vo2013}. For the standard multi-object
system model, the family of GLMB densities is a conjugate prior that
is also closed under the Chapman-Kolmogorov equation \cite{Vo2013}.
Thus, in a dynamic Bayesian inference application, the posterior density
of the point process at each time epoch is a GLMB, which can be tractably
computed using the algorithm developed in \cite{Vo2014}. Recent applications
to problems in multi-object systems such as multiple target tracking,
sensor management, and simultaneous localization and mapping \cite{Deusch2015,Reuter2014,Beard2015,Gostar2014,Papi2015},
suggest that the GLMB is a versatile model that offers good trade-offs
between tractability and fidelity.

In this paper, we derive a closed form void-probability functional
and Cauchy-Schwarz divergence for GLMBs\footnote{Preliminary work on the Cauchy-Schwarz divergence for GLMBs, with
an application to trajectory planning for bearings-only tracking,
has appeared in the authors' conference paper \cite{Beard2015a}.}. Given the theoretical significance of these results in point process
theory, their derivations are remarkably simple. Moreover, these results
provide additional tools to tackle more complex problems in multi-object
systems. To demonstrate the use of both results, we develop a principled
solution to an observer trajectory optimization problem for multi-target
tracking, where the goal is to obtain the most accurate estimates
for an unknown and time-varying number of targets, whilst maintaining
a safe distance from any of the targets. Using GLMBs to model the
collection of targets, we formulate the problem as a constrained partially
observed Markov decision process, with a Cauchy-Schwarz divergence
based reward function and a void-probability based constraint, both
of which can be evaluated analytically.

The paper is organized as follows. In Section \ref{s:Background}
we provide some background on random finite sets, void probabilities
and the Cauchy-Schwarz divergence. In Section \ref{s:GLMB} we describe
the GLMB point process model and its properties, including analytic
expressions for the void probability and Cauchy-Schwarz divergence.
In Section \ref{s:Sensor_Management} we present an application of
the new results, by developing a solution to a sensor management problem
for dynamic Bayesian inference of a point process. Finally, we make
some concluding remarks in Section \ref{s:Conclusion}.

\section{Background\label{s:Background}}

In this work, we consider a state space $\mathbb{X}\subseteq\mathbb{R}^{d}$,
and adopt the inner product notation $\left\langle f,g\right\rangle \triangleq\int f\left(x\right)g\left(x\right)dx$;
the $L^{2}$-norm notation $\left\Vert f\right\Vert \triangleq\sqrt{\left\langle f,f\right\rangle }$;
the multi-object exponential notation $f^{X}\triangleq\prod_{x\in X}f\left(x\right)$,
with $f^{\emptyset}=1$ by convention. Given a set $S$, $1_{S}\left(\cdot\right)$
denotes the indicator function of $S$, $\mathcal{F}\left(S\right)$
denotes the class of finite subsets of $S$, and $S^{i}$ denotes
the $i^{\text{th}}$-fold Cartesian product of $S$, with the convention
$S^{0}=\left\{ \emptyset\right\} $. The cardinality (or number of
elements) of a finite set $X$ is denoted by $\left|X\right|$. The
Kroneker delta function is denoted by $\delta_{n}\left[\cdot\right]$,
which is 1 if the argument is equal to $n$, and 0 otherwise.

\subsection{Random Finite Sets\label{s:RFS}}

Point process theory, in general, is concerned with \emph{random counting
measures}. Our results are restricted to the simple-finite point processes,
which can be regarded as \emph{random finite sets} (RFSs) \cite{Baddeley2007}.
Hence, for compactness we omit the prefix ``simple-finite\textquotedblright{}
and use the terms point process and RFS interchangably. For an introduction
to the subject we refer the reader to the article \cite{Baddeley2007},
and for detailed treatments, textbooks such as \cite{Daley1988,Stoyan1995,Moller2004}.

A \emph{random finite set} (RFS) $X$, defined on the space $\mathbb{X}$,
is a random variable taking values in $\mathcal{F}\left(\mathbb{X}\right)$,
i.e. a finite set-valued random variable. Both the number of elements
and the value of the elements of an RFS are random. There are several
constructs for specifying the probability law of an RFS. The most
convenient of these for the exposition of the two main results of
this paper is the \emph{belief }(or\emph{ containment})\emph{ functional}
$B$, given by
\begin{equation}
B\left(S\right)=\Pr\left(X\subseteq S\right)\label{e:Belief_Functional}
\end{equation}
for any (closed) $S$ $\subseteq$ $\mathbb{X}$ \cite{Stoyan1995,Molchanov2006,Mahler2014,Matheron1975}.
In fact, the belief functional uniquely determines the probability
law of a random closed set (and hence an RFS) via Choquet's capacity
theorem.

\subsection{Void Probability Functional\label{s:Void_Probability_Functional}}

The void probability of an RFS $X$ on a (compact) subset $S\subseteq\mathbb{X}$
is the probability that $S$ contains no points of $X$, i.e. $\Pr\left(X\cap S=\emptyset\right)$,
or equivalently the probability that $X$ is contained in the complement
of $S$, i.e. $\Pr\left(X\subseteq\mathbb{X}-S\right)$ \cite{Stoyan1995,Molchanov2006}.
Thus, in terms of the belief functional the \emph{void probability}
(or \emph{avoidance}) \emph{functional} $Q$ is given by
\begin{equation}
Q\left(S\right)=B\left(\mathbb{X}-S\right).\label{e:Void_Probability}
\end{equation}
As a consequence of Choquet's capacity theorem \cite{Molchanov2006,Matheron1975},
the probability law of an RFS is uniquely defined by the void probability
functional. Rényi also established, using a different line of argument,
that the law of a simple point process is uniquely determined by the
void probabilities on the bounded Borel sets of $\mathbb{X}$ \cite{Daley1988,Stoyan1995,Moller2004}.

The void probability functional is a more intuitive descriptor of
an RFS than its probability distribution. Void probabilities at different
regions provide a sense of\textcolor{blue}{{} }how the number and locations
of the points in an RFS are distributed across the state space. The
concept of void probability is also directly applicable to the more
general class of random closed sets. Indeed, Choquet's capacity theorem
implies that the void probabilities uniquely determine the probability
law of a random closed set \cite{Stoyan1995,Molchanov2006,Matheron1975}.
Note that the void probability on a given region for the union of
two independent RFSs is simply the product of their individual void
probabilities.

\subsection{Cauchy-Schwarz Divergence\label{s:Cauchy-Schwarz_Divergence}}

Geometrically, the Cauchy-Schwarz divergence determines the information
``difference\textquotedblright{} between random variables, from the
angle subtended by their probability densities \cite{Jenssen2005,Jenssen2006}.
Algebraically, it is based on the Cauchy-Schwarz inequality for the
inner product between the probability densities of the random variables.

Since the belief functional is not a measure, the standard notion
of density as a Radon-Nikodym derivative is not applicable. Nevertheless,
an alternative notion of density can be defined via Mahler's set calculus
\cite{Mahler2003,Mahler2014}. The \emph{belief density} of an RFS
$X$ is a non-negative function $\pi$ on $\mathcal{F}\left(\mathbb{X}\right)$
such that for any $S$ $\subseteq$ $\mathbb{X}$,
\begin{equation}
B\left(S\right)=\int_{S}\pi\left(X\right)\delta X,\label{e:Belief_Density}
\end{equation}
where the integral above is Mahler's \emph{set integral} defined by
\cite{Mahler2003,Mahler2014}
\begin{equation}
\int_{S}\pi\left(X\right)\delta X=\sum_{i=0}^{\infty}\frac{1}{i!}\int_{S^{i}}\pi\left(\left\{ x_{1},...,x_{i}\right\} \right)d\left(x_{1},...,x_{i}\right)\label{e:Set_Integral}
\end{equation}
(note that since $S^{0}=\left\{ \emptyset\right\} $, the integral
over $S^{0}$ is simply $\pi\left(\emptyset\right)$). That is, the
set integral of the belief density $\pi$ over a region $S$, yields
the probability that $X$ is contained in $S$. Note that $\pi\left(X\right)$
has dimension $K^{-\left|X\right|}$, where $K$ denotes the unit
of hyper-volume on $\mathbb{X}$.

While the belief density $\pi$ is not a probability density, the
dimensionless function on $\mathcal{F}\left(\mathbb{X}\right)$ defined
by $\pi\left(X\right)K^{\left|X\right|}$ is indeed a probability
density with respect to the measure $\mu$ given by \cite{Vo2005}
\begin{equation}
\mu\left(\mathcal{T}\right)=\sum_{i=0}^{\infty}\frac{1}{i!K^{i}}\int_{\mathcal{X}^{i}}1_{\mathcal{T}}\left(\left\{ x_{1},...,x_{i}\right\} \right)d\left(x_{1},...,x_{i}\right)\label{e:Reference_Measure}
\end{equation}
for any measurable $\mathcal{T}$ $\subseteq$ $\mathcal{F}\left(\mathbb{X}\right)$.

In \cite{Hoang2015}, the Cauchy-Schwarz divergence was extended to
random finite sets via the inner product of their probability densities
relative to the reference measure $\mu$. Using the relationship between
probablity density and belief density, the Cauchy-Schwarz divergence
between two RFSs, with respective belief densities $\phi$ and $\varphi$,
is given by
\begin{equation}
D_{CS}\!\left(\phi,\varphi\right)=-\ln\frac{\int K^{\left|X\right|}\phi\!\left(\!X\!\right)\varphi\!\left(\!X\!\right)\!\delta X}{\sqrt{\!\int\!K^{\left|X\right|}\phi^{2}\!\left(\!X\!\right)\!\delta X\int\!K^{\left|X\right|}\varphi^{2}\!\left(\!X\!\right)\!\delta X}}.\label{e:CS_Divergence}
\end{equation}
Note that $D_{CS}\left(\phi,\varphi\right)$ is invariant to the unit
of hyper-volume $K$ (using the same line of arguments as in \cite[Section III-A]{Hoang2015}).

\subsection{Poisson Point Process\label{s:PPP}}

The \emph{intensity function} of an RFS $X$, is a non-negative function
$v$ (on $\mathbb{X}$) whose integral over any (Borel) $S$ $\subseteq\mathbb{X}$
gives the expected number of elements of the RFS that are in $S$
\cite{Daley1988,Stoyan1995,Moller2004}, i.e. 
\begin{equation}
\mathbb{E}\left[\left\vert X\cap S\right\vert \right]=\left\langle 1_{S},v\right\rangle .\label{e:PHD}
\end{equation}
Since $\left\langle 1_{S},v\right\rangle $ is the expected number
of points of $X$ in the region $S$, the intensity value $v\left(x\right)$
can be interpreted as the instantaneous expected number of points
per unit hyper-volume at $x$. Thus, in general, $v\left(x\right)$
is not dimensionless, but has units of $K^{-1}$. The intensity function
is the first moment of an RFS, and can be computed from the belief
density $\pi$ by \cite{Mahler2003,Mahler2014}
\begin{equation}
v\left(x\right)=\int\pi\left(\left\{ x\right\} \cup X\right)\delta X.\label{e:PHD 2}
\end{equation}

A \emph{Poisson} point process is completely characterized by its
intensity function $v$. The cardinality of a Poisson point process
is Poisson distributed with mean $\left\langle 1,v\right\rangle $,
and conditional on the cardinality, its elements are independently
and identically distributed (i.i.d.) according to the probability
density $v\left(\cdot\right)/\left\langle 1,v\right\rangle $ \cite{Daley1988,Stoyan1995,Moller2004}.
It is implicit that $\left\langle 1,v\right\rangle $ is finite, since
we only consider simple-finite point processes.

The void probability functional and belief density of a Poisson point
process with intensity function $v$ are given, respectively, by \cite{Stoyan1995,Moller2004,Mahler2003,Mahler2014}
\begin{eqnarray}
Q\left(S\right) & = & e^{-\left\langle 1_{S},v\right\rangle },\label{e:PPP_Void_Probability}\\
\pi\left(X\right) & = & e^{-\left\langle 1,v\right\rangle }v^{X}.\label{e:PPP_Belief_Density}
\end{eqnarray}
Moreover, the Cauchy-Schwarz divergence between two Poisson point
processes is given by half the squared $L^{2}$-distance between their
intensity functions \cite{Hoang2015}. As a consequence, the Bhattacharyya
distance between the probability distributions of two Poisson point
processes is the squared Hellinger distance between their respective
intensity measures. For Gaussian mixture intensity functions, the
Cauchy-Schwarz divergence can be evaluated analytically. These results
were also extended to mixtures of Poisson point processes \cite{Hoang2015}.

\section{Generalized Labeled Multi-Bernoulli\label{s:GLMB}}

The Poisson point process is endowed with many elegant mathematical
properties \cite{Daley1988,Stoyan1995,Kingman1993}, including analytic
void probabilities and Cauchy-Schwarz divergence, but it is rather
simplistic for many practical problems. Bayesian inference of hidden
(possibly dynamic) point processes from observed data is a fundamental
problem that arises in multi-object systems, with applications spanning
several disciplines. For most data models, the posterior distributions
of the underlying point processes are not Poisson \cite{Mahler2003,Mahler2014}.
Although Poisson approximations, such as probability hypothesis density
(PHD) filters, are numerically attractive \cite{Mahler2003}, the
Poisson model can neither capture the dependence between the points,
nor permit the inference of the trajectories of the points over time.

The generalized labelled multi-Bernoulli (GLMB) is a class of tractable
models for on-line Bayesian inference that alleviates the limitations
of the Poisson model \cite{Vo2013,Vo2014}. Although sophisticated
models in the spatial point process literature such as Cox, Neyman-Scott,
Gauss-Poisson, Markov (or Gibbs) \cite{Daley1988,Stoyan1995,Moller2004},
are able to accommodate interactions such as repulsion, attraction
or clustering, they cannot capture exactly the general inter-point
dependencies in the posterior distribution that transpires through
the data. In other words, they are not conjugate with respect to the
standard multi-object measurement likelihood function. Moreover, these
models are neither amenable to on-line computation, nor to the inference
of trajectories.

In this section, we revisit the GLMB model \cite{Vo2013} and some
of its analytical properties. In addition, we present closed form
expressions for the void probability functional and the Cauchy-Schwarz
divergence for the GLMB.

\subsection{Labeled RFS\label{s:Labeled_RFS}}

Let $\mathbb{L}$ be a discrete space, and $\mathcal{L}:\mathbb{X}\mathcal{\times}\mathbb{L}\rightarrow\mathbb{L}$
be the projection defined by $\mathcal{L}\left(x,\ell\right)=\ell$.
Then $\mathcal{L}\left(\mathbf{x}\right)$ is called the label of
the point $\mathbf{x}\in\mathbb{X}\mathcal{\times}\mathbb{L}$, and
a finite subset $\mathbf{X}$ of $\mathbb{X}\mathcal{\times}\mathbb{L}$
is said to have \emph{distinct labels} if and only if $\mathbf{X}$
and its label set $\mathcal{L}\left(\mathbf{X}\right)=\left\{ \mathcal{L}\left(\mathbf{x}\right):\mathbf{x}\in\mathbf{X}\right\} $
have the same cardinality.

A \emph{labeled RFS} is a marked point process with state space $\mathbb{X}$
and mark space $\mathbb{L}$ such that each realization has distinct
labels \cite{Vo2013}. In dynamic Bayesian inference, the posterior
distribution of the underlying point process is computed recursively
in time as data arrives, and the distinct labels provide a means of
identifying the trajectories of individual points. A trajectory is
defined as a time-sequence of points with the same label. The distinct
label property ensures that, at any given time instant, no two points
can share the same label, and hence no two trajectories can share
any common points.

The unlabeled version of a labeled RFS is its projection from $\mathbb{X}\mathcal{\times}\mathbb{L}$
into $\mathbb{X}$, and is obtained by simply discarding the labels.
The cardinality distributions of a labeled RFS and its unlabeled counterpart
are identical \cite{Vo2013}. However, the intensity function $v\left(\cdot,\cdot\right)$
(defined on $\mathbb{X}\times\mathbb{L}$) of a labeled RFS is related
to its unlabeled counterpart $v\left(\cdot\right)$ (defined on $\mathbb{X}$)
by \cite{Vo2013}
\begin{equation}
v\left(x\right)=\sum_{\ell\in\mathbb{L}}v\left(x,\ell\right).
\end{equation}
For the rest of the paper, points are represented by lowercase (e.g.
$x$, $\mathbf{x}$), while point patterns (or finite sets of points)
are represented by uppercase (e.g. $X$, $\mathbf{X}$). Symbols for
labeled points, labeled point patterns, and their distributions are
bolded (e.g. $\mathbf{x}$, $\mathbf{X}$, $\boldsymbol{\pi}$) to
distinguish them from unlabeled ones, and spaces are represented by
blackboard bold (e.g. $\mathbb{X}$, $\mathbb{Z}$, $\mathbb{L}$).

\subsection{GLMBs and their Properties\label{s:GLMB_Properties}}

A GLMB is a labeled RFS with belief density on $\mathcal{F}\left(\mathbb{X}\mathcal{\times}\mathbb{L}\right)$
of the form
\begin{equation}
\boldsymbol{\pi}\left(\mathbf{X}\right)=\Delta\left(\mathbf{X}\right)\sum_{c\in\mathbb{C}}w^{\left(c\right)}\left(\mathcal{L}\left(\mathbf{X}\right)\right)\left[p^{\left(c\right)}\right]^{\boldsymbol{X}},\label{e:GLMB}
\end{equation}
where $\Delta\left(\mathbf{X}\right)\triangleq\delta_{\left|\mathbf{X}\right|}\left(\left|\mathcal{L}\left(\mathbf{X}\right)\right|\right)$
is the \emph{distinct label indicator}, $\mathbb{C}$ is a discrete
and finite index set, each $p^{\left(c\right)}\left(\cdot,\ell\right)$
is a probability density on $\mathbb{X}$, and each $w^{\left(c\right)}\left(L\right)$
is non-negative with 
\begin{equation}
\sum_{L\subseteq\mathbb{L}}\sum_{c\in\mathbb{C}}w^{\left(c\right)}\left(L\right)=1.
\end{equation}

By convention, $p^{\left(c\right)}\left(x,\ell\right)$ are measured
in units of $K^{-1}$, and consequently, $\boldsymbol{\pi}\left(\boldsymbol{X}\right)$
has units of $K^{-\left|\boldsymbol{X}\right|}$. The belief density
\eqref{e:GLMB} is a mixture of multi-object exponentials, with each
component consisting of a weight $w^{\left(c\right)}\left(\mathcal{L}\left(\mathbf{X}\right)\right)$
that depends only on the labels of $\mathbf{X}$, and a multi-object
exponential $\left[p^{\left(c\right)}\right]^{\boldsymbol{X}}$. Such
a structure provides the flexibility for the GLMB to capture the dependence
between points that transpires via the data, and also admits a number
of convenient analytical properties, which are summarised as follows.
\begin{itemize}
\item For the standard multi-object system model that accounts for thinning,
Markov shifts and superposition, the GLMB family is a conjugate prior,
and is also closed under the Chapman-Kolmogorov equation \cite{Vo2013}.
\item The GLMB density can be approximated to any $L_{1}$-norm error by
truncating components \cite{Vo2014}. More precisely, let us explicitly
denote the dependence on the index set $\mathbb{C}$ of a (possibly
unnormalized) GLMB density by
\begin{equation}
\mathbf{f}_{\mathbb{C}}\mathbf{\left(X\right)}=\Delta\left(\mathbf{X}\right)\sum\limits _{c\in\mathbb{C}}w^{\left(c\right)}\left(\mathcal{L}\left(\mathbf{X}\right)\right)\left[p^{\left(c\right)}\right]^{\mathbf{X}}
\end{equation}
and let $\left\Vert \mathbf{f}\right\Vert _{1}\triangleq\int\left|\mathbf{f}\left(\mathbf{X}\right)\right|\delta\mathbf{X}$
denote the $L_{1}$-norm of $\mathbf{f:\mathcal{F}}\left(\mathbb{X}\mathcal{\times}\mathbb{L}\right)\rightarrow\mathbb{R}$.
If $\mathbb{D}\subseteq\mathbb{C}$ then
\begin{align}
\left|\left|\mathbf{f}_{\mathbb{C}}-\mathbf{f}_{\mathbb{\mathbb{D}}}\right|\right|_{1} & \mathbf{=}\sum\limits _{c\in\mathbb{C}-\mathbb{D}}\sum\limits _{L\subseteq\mathbb{L}}w^{\left(c\right)}\left(L\right).
\end{align}

\item The cardinality distribution and intensity function of a GLMB are
respectively given by
\begin{align}
\Pr\left(\left\vert X\right\vert \text{=}n\right) & =\sum_{c\in\mathbb{C}}\sum_{L\subseteq\mathbb{L}}\delta_{n}\left[\left\vert L\right\vert \right]w^{\left(c\right)}\left(L\right),\label{e:GLMB_Cardinality}\\
v\left(x,\ell\right) & =\sum_{c\in\mathbb{C}}p^{\left(c\right)}\left(x,\ell\right)\sum_{L\subseteq\mathbb{L}}1_{L}\left(\ell\right)w^{\left(c\right)}\left(L\right).\label{e:GLMB_Intensity}
\end{align}

\item The GLMB is flexible enough to approximate any labeled RFS density,
by matching the intensity function and cardinality distribution. Furthermore,
there is a simple closed form that mimimizes the Kullback-Leibler
divergence between the labelled RFS density and its GLMB approximation
\cite{Papi2015a}.
\end{itemize}
As shown above, the GLMB family possesses some useful analytical properties.
There is also an elegant characterisation of the GLMB using the probability
generating functional (p.g.fl.) by Mahler \cite{Mahler2014}. In the
following subsection, we derive two additional properties of the GLMB,
which have some potentially useful applications.

\subsection{Void Probability Functional\label{s:GLMB_Void_Probability}}
\begin{prop}
For a GLMB with belief density $\boldsymbol{\pi}$ of the form \eqref{e:GLMB},
the void probability functional is given by
\begin{align}
Q_{\pi}\left(S\right) & =\sum_{L\subseteq\mathbb{L}}\sum_{c\in\mathbb{C}}w^{\left(c\right)}\left(L\right)\prod_{\ell\in L}\left\langle 1-1_{S},p^{\left(c\right)}\left(\cdot,\ell\right)\right\rangle .\label{e:GLMB_Void_Probability}
\end{align}
\end{prop}
\begin{IEEEproof}
Using \eqref{e:Void_Probability} and \eqref{e:Belief_Density}, the
void probability functional can be expressed as
\begin{align}
Q_{\boldsymbol{\pi}}\left(S\right) & =\int_{\mathbb{X}-S}\boldsymbol{\pi}\left(X\right)\delta X\\
 & =\int_{\mathbb{X}-S}\!\Delta\!\left(\boldsymbol{X}\right)\sum_{c\in\mathbb{C}}w^{\left(c\right)}\!\left(\mathcal{L}\left(\boldsymbol{X}\right)\right)\!\left[p^{\left(c\right)}\left(\cdot\right)\right]^{\boldsymbol{X}}\!\delta\boldsymbol{X}.
\end{align}
Applying Lemma 3 from \cite{Vo2013}, yields the result
\begin{align}
Q_{\boldsymbol{\pi}}\left(S\right) & =\sum_{L\subseteq\mathbb{L}}\sum_{c\in\mathbb{C}}w^{\left(c\right)}\left(L\right)\left[\int_{\mathbb{X}-S}p^{\left(c\right)}\left(x,\cdot\right)dx\right]^{L}\\
 & =\sum_{L\subseteq\mathbb{L}}\sum_{c\in\mathbb{C}}w^{\left(c\right)}\left(L\right)\prod_{\ell\in L}\left\langle 1-1_{S},p^{\left(c\right)}\left(\cdot,\ell\right)\right\rangle .\label{e:GLMB_Void_Probability-1}
\end{align}

\end{IEEEproof}
In simple cases, the inner product $\left\langle 1-1_{S},p^{\left(c\right)}\left(\cdot,\ell\right)\right\rangle $
may be computable in closed form, in which case the void probability
has an exact analytical solution. However, in more general cases,
a closed form may not exist, and it must therefore be computed using
numerical methods such as cubature or Monte Carlo integration. This
yields an approximation to the true void probability for the GLMB.

In general, the computational complexity is $\mathcal{O}\left(N+M\right)$,
where $N$ is the number of pairs $\left(c,L\right)\in\mathbb{C}\times\mathcal{F}\left(\mathbb{L}\right)$
such that $w^{\left(c\right)}\left(L\right)\neq0$, and $M$ is the
number of unique single-object densities $p^{\left(c\right)}\left(\cdot,\ell\right)$.
In many applications, the GLMB may contain a large number of single-object
densities which are common across many elements of the sum, in which
case the inner product only needs to be computed once for each unique
single-object density.

The analytic expression for the GLMB void probability functional is
of theoretical interest in itself, since it provides an alternative
means of completely specifying a GLMB point process. However, it also
holds significant practical interest, since it can be used to compute
statistics that can conceivably be applied in a wide range of real-world
problems.

In multi-object estimation and control, the GLMB void probability
functional could supply useful information that can be applied in
tasks such as trajectory planning (e.g. for collision avoidance),
sensor management (e.g. focussing sensor resources on regions where
targets are likely to be present), or the provision of situational
awareness (e.g. advance warning of possible collisions between objects).
The GLMB is a flexible model which has been used to develop algorithms
for target tracking \cite{Vo2013,Vo2014,Reuter2014a} and simultaneous
localization and mapping \cite{Deusch2015}. Indeed, the GLMB has
been applied in autonomous vehicle systems \cite{Reuter2014a}, where
trajectory planning and situational awareness for collision avoidance
are paramount. It has also been applied to the tracking of orbital
space debris \cite{Jones2014,Jones2015}, for which the scheduling
and management of observation equipment is a significant issue, as
well as the planning of satellite trajectories to minimize the probability
of collision with tracked debris.

\subsection{Cauchy-Schwarz Divergence\label{s:GLMB_Cauchy-Schwarz}}

Using the definition in equation (\ref{e:CS_Divergence}), we show
that the Cauchy-Schwarz divergence between two GLMBs can be expressed
in closed form.
\begin{prop}
\label{p:GLMB_CSD}For two GLMBs with belief densities
\begin{align}
\boldsymbol{\phi}\left(\boldsymbol{X}\right) & =\Delta\left(\boldsymbol{X}\right)\sum_{c\in\mathbb{C}}w_{\boldsymbol{\phi}}^{\left(c\right)}\left(\mathcal{L}\left(\boldsymbol{X}\right)\right)\left[p_{\boldsymbol{\phi}}^{\left(c\right)}\left(\cdot\right)\right]^{\boldsymbol{X}},\label{e:GLMB_1}\\
\boldsymbol{\psi}\left(\boldsymbol{X}\right) & =\Delta\left(\boldsymbol{X}\right)\sum_{d\in\mathbb{D}}w_{\boldsymbol{\psi}}^{\left(d\right)}\left(\mathcal{L}\left(\boldsymbol{X}\right)\right)\left[p_{\boldsymbol{\psi}}^{\left(d\right)}\left(\cdot\right)\right]^{\boldsymbol{X}},\label{e:GLMB_2}
\end{align}
where both $p_{\boldsymbol{\phi}}^{\left(c\right)}$ and $p_{\boldsymbol{\psi}}^{\left(d\right)}$
are measured in units of $K^{-1}$, the Cauchy-Schwarz divergence
between $\boldsymbol{\phi}$ and $\boldsymbol{\psi}$ is given by
\begin{equation}
D_{CS}\left(\boldsymbol{\phi},\boldsymbol{\psi}\right)=-\ln\left(\frac{\left\langle \boldsymbol{\phi},\boldsymbol{\psi}\right\rangle _{K}}{\sqrt{\left\langle \boldsymbol{\phi},\boldsymbol{\phi}\right\rangle _{K}\left\langle \boldsymbol{\psi},\boldsymbol{\psi}\right\rangle _{K}}}\right),\label{e:GLMB_Cauchy-Schwarz}
\end{equation}
where
\begin{align}
\left\langle \boldsymbol{\phi},\boldsymbol{\psi}\right\rangle _{K} & =\sum_{L\subseteq\mathbb{L}}\sum_{\substack{c\in\mathbb{C}\\
d\in\mathbb{D}
}
}w_{\boldsymbol{\phi}}^{\left(c\right)}\!\left(\!L\!\right)w_{\boldsymbol{\psi}}^{\left(d\right)}\!\left(\!L\!\right)\!\prod_{\ell\in L}\!K\!\left\langle \!p_{\boldsymbol{\phi}}^{\left(c\right)}\!\left(\cdot,\ell\right),p_{\boldsymbol{\psi}}^{\left(d\right)}\!\left(\cdot,\ell\right)\!\right\rangle \nonumber \\
\left\langle \boldsymbol{\phi},\boldsymbol{\phi}\right\rangle _{K} & =\sum_{L\subseteq\mathbb{L}}\sum_{\substack{c\in\mathbb{C}\\
d\in\mathbb{C}
}
}w_{\boldsymbol{\phi}}^{\left(c\right)}\!\left(\!L\!\right)w_{\boldsymbol{\phi}}^{\left(d\right)}\!\left(\!L\!\right)\!\prod_{\ell\in L}\!K\!\left\langle \!p_{\boldsymbol{\phi}}^{\left(c\right)}\!\left(\cdot,\ell\right),p_{\boldsymbol{\phi}}^{\left(d\right)}\!\left(\cdot,\ell\right)\!\right\rangle \nonumber \\
\left\langle \boldsymbol{\psi},\boldsymbol{\psi}\right\rangle _{K} & =\sum_{L\subseteq\mathbb{L}}\sum_{\substack{c\in\mathbb{D}\\
d\in\mathbb{D}
}
}w_{\boldsymbol{\psi}}^{\left(c\right)}\!\left(\!L\!\right)w_{\boldsymbol{\psi}}^{\left(d\right)}\!\left(\!L\!\right)\!\prod_{\ell\in L}\!K\!\left\langle \!p_{\boldsymbol{\psi}}^{\left(c\right)}\!\left(\cdot,\ell\right),p_{\boldsymbol{\psi}}^{\left(d\right)}\!\left(\cdot,\ell\right)\!\right\rangle \label{e:GLMB_Inner_Products}
\end{align}
\end{prop}
\begin{IEEEproof}
If $\boldsymbol{\phi}\left(\boldsymbol{X}\right)$ and $\boldsymbol{\psi}\left(\boldsymbol{X}\right)$
are two GLMBs defined by \eqref{e:GLMB_1} and \eqref{e:GLMB_2},
the inner product is given by
\begin{align*}
\left\langle \boldsymbol{\phi},\boldsymbol{\psi}\right\rangle _{K} & =\int K^{\left|\boldsymbol{X}\right|}\boldsymbol{\phi}\left(\boldsymbol{X}\right)\boldsymbol{\psi}\left(\boldsymbol{X}\right)\delta\boldsymbol{X}\\
 & =\int K^{\left|\boldsymbol{X}\right|}\Delta\left(\boldsymbol{X}\right)\sum_{c\in\mathbb{C}}w_{\boldsymbol{\phi}}^{\left(c\right)}\left(\mathcal{L}\left(\boldsymbol{X}\right)\right)\left[p_{\boldsymbol{\phi}}^{\left(c\right)}\left(\cdot\right)\right]^{\boldsymbol{X}}\\
 & \qquad\times\sum_{d\in\mathbb{D}}w_{\boldsymbol{\psi}}^{\left(d\right)}\left(\mathcal{L}\left(\boldsymbol{X}\right)\right)\left[p_{\boldsymbol{\psi}}^{\left(d\right)}\left(\cdot\right)\right]^{\boldsymbol{X}}\delta\boldsymbol{X}\\
 & =\int\Delta\left(\boldsymbol{X}\right)\sum_{c\in\mathbb{C}}\sum_{d\in\mathbb{D}}w_{\boldsymbol{\phi}}^{\left(c\right)}\left(\mathcal{L}\left(\boldsymbol{X}\right)\right)w_{\boldsymbol{\psi}}^{\left(d\right)}\left(\mathcal{L}\left(\boldsymbol{X}\right)\right)\\
 & \qquad\times\left[Kp_{\boldsymbol{\phi}}^{\left(c\right)}\left(\cdot\right)p_{\boldsymbol{\psi}}^{\left(d\right)}\left(\cdot\right)\right]^{\boldsymbol{X}}\delta\boldsymbol{X}.
\end{align*}
Making use of Lemma 3 in \cite{Vo2013}, this becomes
\begin{align*}
\left\langle \boldsymbol{\phi},\boldsymbol{\psi}\right\rangle _{K} & =\!\sum_{L\subseteq\mathbb{L}}\sum_{\substack{c\in\mathbb{C}\\
d\in\mathbb{D}
}
}\!w_{\boldsymbol{\phi}}^{\left(c\right)}\!\left(\!L\!\right)w_{\boldsymbol{\psi}}^{\left(d\right)}\!\left(\!L\!\right)\!\left[\!K\!\!\int\!p_{\boldsymbol{\phi}}^{\left(c\right)}\!\left(x,\cdot\right)p_{\boldsymbol{\psi}}^{\left(d\right)}\!\left(x,\cdot\right)dx\!\right]^{L}\\
 & =\!\sum_{L\subseteq\mathbb{L}}\sum_{\substack{c\in\mathbb{C}\\
d\in\mathbb{D}
}
}w_{\boldsymbol{\phi}}^{\left(c\right)}\!\left(\!L\!\right)w_{\boldsymbol{\psi}}^{\left(d\right)}\!\left(\!L\!\right)\!\prod_{\ell\in L}\!K\!\left\langle \!p_{\boldsymbol{\phi}}^{\left(c\right)}\!\left(\cdot,\ell\right),p_{\boldsymbol{\psi}}^{\left(d\right)}\!\left(\cdot,\ell\right)\!\right\rangle 
\end{align*}
and similarly for $\left\langle \boldsymbol{\phi},\boldsymbol{\phi}\right\rangle _{K}$
and $\left\langle \boldsymbol{\psi},\boldsymbol{\psi}\right\rangle _{K}$.
Substituting these into \eqref{e:CS_Divergence}, yields the result
\eqref{e:GLMB_Cauchy-Schwarz}-\eqref{e:GLMB_Inner_Products}.
\end{IEEEproof}
In cases where the inner product between two single-object densities
$p_{\boldsymbol{\phi}}^{\left(c\right)}$ and $p_{\boldsymbol{\psi}}^{\left(d\right)}$
of the GLMBs has an analytical solution, then the Cauchy-Schwarz divergence
between the two GLMBs can also be evaluated analytically. However,
where this is not possible, numerical approximations can be used to
evaluate the inner products. The common case in which the single-object
densities are Gaussian mixtures, does admit an analytical solution,
as established in the following proposition.
\begin{prop}
Let $\boldsymbol{\phi}$ and $\boldsymbol{\psi}$ be two GLMBs of
the form \eqref{e:GLMB_1} and \eqref{e:GLMB_2} in which the single-object
densities are Gaussian mixtures, i.e.
\begin{align}
p_{\boldsymbol{\phi}}^{\left(c\right)}\left(x,\ell\right) & =\sum_{i=1}^{N_{\boldsymbol{\phi}}^{\left(c\right)}}\omega_{\boldsymbol{\phi},i}^{\left(c\right)}\left(\ell\right)\mathcal{N}\left(x;m_{\boldsymbol{\phi},i}^{\left(c\right)}\left(\ell\right),P_{\boldsymbol{\phi},i}^{\left(c\right)}\left(\ell\right)\right),\label{e:Single_Object_1}\\
p_{\boldsymbol{\psi}}^{\left(d\right)}\left(x,\ell\right) & =\sum_{i=1}^{N_{\boldsymbol{\psi}}^{\left(d\right)}}\omega_{\boldsymbol{\psi},i}^{\left(d\right)}\left(\ell\right)\mathcal{N}\left(x;m_{\boldsymbol{\psi},i}^{\left(d\right)}\left(\ell\right),P_{\boldsymbol{\psi},i}^{\left(d\right)}\left(\ell\right)\right),\label{e:Single_Object_2}
\end{align}
where both $p_{\boldsymbol{\phi}}^{\left(c\right)}$ and $p_{\boldsymbol{\psi}}^{\left(d\right)}$
are measured in units of $K^{-1}$. The Cauchy-Schwarz divergence
between $\boldsymbol{\phi}$ and $\boldsymbol{\psi}$ can be expressed
in analytical form, since $\left\langle \boldsymbol{\phi},\boldsymbol{\psi}\right\rangle _{K}$
in \eqref{e:GLMB_Inner_Products} reduces to
\begin{equation}
\left\langle \boldsymbol{\phi},\boldsymbol{\psi}\right\rangle _{K}=\sum_{L\subseteq\mathbb{L}}\sum_{c\in\mathbb{C}}\sum_{d\in\mathbb{D}}w_{\boldsymbol{\phi}}^{\left(c\right)}\left(L\right)w_{\boldsymbol{\psi}}^{\left(d\right)}\left(L\right)\left[\gamma_{\phi,\psi}\right]^{L}\label{e:GM_GLMB_Inner_Prod}
\end{equation}
where
\begin{align}
\gamma_{\boldsymbol{\phi},\boldsymbol{\psi}}\left(\ell\right) & =\sum_{i=1}^{N_{\boldsymbol{\phi}}^{\left(c\right)}}\sum_{j=1}^{N_{\boldsymbol{\psi}}^{\left(d\right)}}\omega_{\boldsymbol{\phi},i}^{\left(c\right)}\left(\ell\right)\omega_{\boldsymbol{\psi},j}^{\left(d\right)}\left(\ell\right)\label{e:GM_GLMB_Inner_Prod_Gamma}\\
 & \qquad\times\mathcal{N}\!\left(m_{\boldsymbol{\phi},i}^{\left(c\right)}\left(\ell\right);m_{\boldsymbol{\psi},j}^{\left(d\right)}\left(\ell\right),P_{\boldsymbol{\phi},i}^{\left(c\right)}\left(\ell\right)\!\!+\!P_{\boldsymbol{\psi},j}^{\left(d\right)}\left(\ell\right)\right),\nonumber 
\end{align}
and similarly for $\left\langle \boldsymbol{\phi},\boldsymbol{\phi}\right\rangle _{K}$
and $\left\langle \boldsymbol{\psi},\boldsymbol{\psi}\right\rangle _{K}$.\end{prop}
\begin{IEEEproof}
Substituting \eqref{e:Single_Object_1} and \eqref{e:Single_Object_2}
into the inner product in \eqref{e:GLMB_Inner_Products}, gives 
\begin{align*}
 & \left\langle p_{\boldsymbol{\phi}}^{\left(c\right)}\left(\cdot,\ell\right),p_{\boldsymbol{\psi}}^{\left(d\right)}\left(\cdot,\ell\right)\right\rangle \\
 & \quad=\sum_{i=1}^{N_{\boldsymbol{\phi}}^{\left(c\right)}}\sum_{j=1}^{N_{\boldsymbol{\psi}}^{\left(d\right)}}\omega_{\boldsymbol{\phi},i}^{\left(c\right)}\left(\ell\right)\omega_{\boldsymbol{\psi},j}^{\left(d\right)}\left(\ell\right)\\
 & \qquad\times\int\!\mathcal{N}\!\left(\!x;m_{\boldsymbol{\phi},i}^{\left(c\right)}\!\left(\ell\right),P_{\boldsymbol{\phi},i}^{\left(c\right)}\!\left(\ell\right)\!\right)\!\mathcal{N}\!\left(\!x;m_{\boldsymbol{\psi},j}^{\left(d\right)}\!\left(\ell\right),P_{\boldsymbol{\psi},j}^{\left(d\right)}\!\left(\ell\right)\!\right)dx,
\end{align*}
which is measured in units of $K^{-1}$. Applying the identity for
a product of two Gaussians \cite[pp. 200]{Rasmussen2005}, and multiplying
by $K$, we are left with the unitless quantity
\begin{align}
K\left\langle p_{\boldsymbol{\phi}}^{\left(c\right)}\left(\cdot,\ell\right),p_{\boldsymbol{\psi}}^{\left(d\right)}\left(\cdot,\ell\right)\right\rangle  & =\gamma_{\boldsymbol{\phi},\boldsymbol{\psi}}\left(\ell\right),\label{e:GM_Inner_Prod}
\end{align}
where $\gamma_{\boldsymbol{\phi},\boldsymbol{\psi}}\left(\ell\right)$
is given by \eqref{e:GM_GLMB_Inner_Prod_Gamma}. Substituting \eqref{e:GM_Inner_Prod}
into \eqref{e:GLMB_Inner_Products}, yields the result \eqref{e:GM_GLMB_Inner_Prod}.
\end{IEEEproof}
\textit{Remark:} A GLMB $\boldsymbol{\phi}$ is completely paramterized
by the set 
\[
\left\{ \left(w_{\boldsymbol{\phi}}^{\left(c\right)}\left(L\right),p_{\boldsymbol{\phi}}^{\left(c\right)}\right):\left(c,L\right)\in\mathbb{C}\times\mathcal{F}\left(\mathbb{L}\right)\right\} 
\]
and we refer to each element $\left(w_{\boldsymbol{\phi}}^{\left(c\right)}\left(L\right),p_{\boldsymbol{\phi}}^{\left(c\right)}\right)$
of this set as a component of a GLMB. Due to the nested summations
in \eqref{e:GLMB_Inner_Products} or \eqref{e:GM_GLMB_Inner_Prod},
a naive implementation will have a computational complexity of $\mathcal{O}\left(MN+M^{2}+N^{2}\right)$,
where $M$ and $N$ are the number of components of $\boldsymbol{\phi}$
and $\boldsymbol{\psi}$ with non-zero weights. This leads to the
summations being taken only over pairs of components with matching
label sets. It is therefore possible to reduce computation using associative
data structures to facilitate finding these matching pairs. Although
this does not reduce the worst case complexity, the average will be
significantly better. 

Note that if $w_{\boldsymbol{\phi}}^{\left(c\right)}\left(L\right)=0$
or $w_{\boldsymbol{\psi}}^{\left(d\right)}\left(L\right)=0$ for all
triples$\left(L,c,d\right)\in\mathcal{F}\left(\mathbb{L}\right)\times\mathbb{C}\times\mathbb{D}$,
then $\left\langle \boldsymbol{\phi},\boldsymbol{\psi}\right\rangle _{K}$
will evaluate to zero, (i.e. $\boldsymbol{\phi}$ and $\boldsymbol{\psi}$
are orthogonal) leading to a Cauchy-Schwarz divergence of infinity.
This is an intuitive result as two such GLMBs have no non-zero components
with matching labels.

\section{Application to Sensor Management\label{s:Sensor_Management}}

In this section, we apply the proposed closed form solutions for the
Cauchy-Schwarz divergence and void probabilities to a multi-target
sensor management problem. In most target tracking scenarios, the
sensor may perform various actions that can have a significant impact
on the quality of the observed data, and can therefore influence the
estimation performance of the tracking system. Typically, such actions
might include changes in the position, orientation or motion of sensor
platforms \cite{Singh2007,Tang2005,Grocholsky2003,Oshman1999}, changes
to sensor deployment and utilization \cite{Tharmarasa2007,Hernandez2004},
or altering the sensor operating parameters such as the beam pattern
\cite{Krishnamurthy2001,Kreucher2005}, or transmit waveform \cite{Kershaw1994,Niu2002}.
The control actions affect the information content of the received
data, which in turn affects the system's ability to detect, track,
and identify the targets.

Often, the control decisions are driven by manual intervention, which
provides no guarantee of optimality. The goal of automatic sensor
management is to determine the best control actions, based on some
optimality criteria. This has the potential to improve tracking performance,
by making control decisions in a systematic and optimal manner that
accounts for the prevailing conditions.

\subsection{Problem Statement\label{s:Problem_Statement}}

In this application, the aim is to perform sensor control in the context
of a multi-target tracking system that is based upon the standard
models of multi-object dynamics and multi-object observations. In
the standard multi-object dynamic model, at time $k-1$, each target
$\boldsymbol{x}_{k-1}$ of a multi-object state $\boldsymbol{X}_{k-1}$
generates a set $\boldsymbol{S}_{k|k-1}(\mathbf{x}_{k-1})$ at time
$k$, which is a singleton if the target survives, or an empty set
if the target dies. New targets appearing at time $k$ are represented
by a set $\boldsymbol{B}_{k}$. Thus, the multi-object state $\boldsymbol{X}_{k}$
generated by $\boldsymbol{X}_{k-1}$ is given by the \emph{multi-object
state transition equation} 
\begin{equation}
\boldsymbol{X}_{k}=\bigcup\limits _{\mathbf{x}_{k-1}\in\boldsymbol{X}_{k-1}}\boldsymbol{S}_{k|k-1}(\mathbf{x}_{k-1})\cup\boldsymbol{B}_{k}.\label{e:RFS_Transition}
\end{equation}
This transition equation captures the underlying models of object
motion, births and deaths, more details of which can be found in \cite[Section IV-D]{Vo2013}.

In the standard multi-object observation model, each target $\mathbf{x}_{k}\in\boldsymbol{X}_{k}$
generates a set $D_{k}(\mathbf{x}_{k})$ which is a singleton if the
target is detected, or empty if the target is misdetected. The observation
$Z_{k}$ generated by $\boldsymbol{X}_{k}$ is given by the \emph{multi-object}
\emph{observation equation}
\begin{equation}
Z_{k}=\bigcup\limits _{\mathbf{x}_{k}\in\boldsymbol{X}_{k}}D_{k}(\mathbf{x}_{k})\cup F_{k},\label{e:RFS_Observation}
\end{equation}
where $F_{k}$ is a set of false detections. Note that in general,
the observation will depend on the chosen control action, which we
omit for compactness of notation. In the standard observation model,
the multi-object observation equation captures underlying models of
target detections, measurement noise, and false alarms, and the reader
is referred to \cite[Section IV-C]{Vo2013} for more details. Note
that the measurements do not contain any specific information to associate
them with a particular object, thus the origin of any particular observation
is not certain.

The quality of the observations (i.e. the detection probability and
measurement noise) is dependent on the state of the objects and the
sensor itself, for example, objects that are further away from the
sensor generally have lower probability of detection and higher measurement
noise. For this reason, the control of the sensor can have a significant
influence on the tracking performance.

Here we address the problem of controlling the motion of a single
sensor platform, with the aim of optimizing the tracking performance
under the aforementioned dynamic and observation models. Since the
control actions affect the observation quality, the goal is to design
a scheme which can automatically select control actions that yield
the most `informative' observations. This is a difficult problem due
to the unknown and time-varying number of targets, and the uncertainty
in the multi-object state due to the measurement noise, object detection/misdetection,
and unknown measurement origin.

\subsection{Control Strategy\label{s:Control_Strategy}}

We now proceed to formulate the control problem as a partially observable
Markov decision processes (POMDP) \cite{Bertsekas1995,Monahan1982,Lovejoy1991,Castanon2008}.
In general terms, the elements of a POMDP are as follows.
\begin{enumerate}
\item The system dynamics is a Markov process.
\item The observations follow a known distribution, conditioned on the state
and the sensor control action.
\item The true state of the system is unknown, but we have access to the
posterior probability density function (pdf) of the state conditioned
on past observations.
\item The benefit of performing a given action can be expressed by a reward
function, which characterises the objectives of the control system.
\end{enumerate}
In this case, the dynamics is modelled by \eqref{e:RFS_Transition}
and the observations are modelled by \eqref{e:RFS_Observation}. The
posterior pdf of the system state is modelled as a GLMB of the form
\eqref{e:GLMB}, since it enables tractable estimation of object trajectories
to inform the control strategy. In general, there are two broad categories
of reward function that can be used in a POMDP, namely, `task-based'
and `information-based' reward. Task-based reward functions (see for
example \cite{Gostar2014}) are useful in situations where the control
problem can be formulated in terms of a single well-defined objective.
However, in situations where this does not exist, the information-based
approach is more appropriate, as it strives to capture the information
gain in an overall sense (for example \cite{Ristic2010}). In this
example, the reward function is formulated in terms of an information
divergence between prior and posterior multi-object densities, which
bears a strong relationship to the improvement in estimation accuracy,
since higher information divergence indicates greater information
gain, which leads to more accure track estimates.

It is also possible to place constraints on the control problem, which
is useful in cases where it is foreseeable that certain actions might
result in undesireable side-effects. For example, to guarantee the
safety or covertness of the sensor platform, we might want to ensure
that no targets enter a predefined exclusion region around the sensor.
This can be achieved using a constrained POMDP \cite{Isom2008,Undurti2010},
in which the goal is to find the control action that maximises the
reward function, subject to one or more constraints.

Within the POMDP framework, the most computationally tractable strategy
is to use myopic open-loop feedback control \cite{Bertsekas1995},
with a discrete action space. The term `myopic' means that the algorithm
only decides one control action at a time, rather than planning multiple
actions into the future.

At the time that a control action is performed, we have no knowledge
of the posterior density that would arise from taking that action.
Since this precludes calculation of the true information divergence,
its expectation with respect to all possible future measurements is
taken \cite{Ristic2010,Mahler2003a}. More precicely, let us begin
by defining the following notation
\begin{itemize}
\item $\boldsymbol{\pi}_{k}\left(\cdot|Z_{1:k}\right)$ is the posterior
density at time $k$,
\item $\mathbb{A}_{k}$ is a discrete space of control actions at time $k$,
\item $H$ is the length of the control horizon,
\item $\boldsymbol{\pi}_{k+H}\left(\cdot|Z_{1:k}\right)$ is the predicted
density at time $k+H$ given measurements up to time $k$,
\item $Z_{k+1:k+H}\left(\alpha\right)$ is the collection of measurement
sets that would be observed from times $k+1$ up to $k+H$, if control
action $\alpha\in\mathbb{A}_{k}$ was executed at time $k$,
\item $V_{k}\left(\alpha\right)$ is the exclusion region around the sensor
at time $k$ under control action $\alpha$,
\item $Q_{\pi}\left(S\right)$ is the void probability functional corresponding
to the multi-object density $\pi$ over region $S$,
\item $P_{vmin}$ is the minimum void probability threshold.
\end{itemize}
The optimal control action is given by maximising the expected value
of a reward function $\mathcal{R}_{k+H}\left(\cdot\right)$ over the
space of allowable actions \cite{Ristic2010}, 
\begin{align}
\alpha_{opt} & =\arg\max_{\alpha\in\mathbb{A}_{k}}\mathbb{E}\left[\mathcal{R}_{k+H}\left(\alpha\right)\right],\label{e:Objective}
\end{align}
subject to the constraint
\begin{equation}
\min_{i\in\left\{ 1,\dots,H\right\} }\left[Q_{\boldsymbol{\pi}_{k+i}\left(\boldsymbol{X}|Z_{1:k}\right)}\left(V_{i}\left(\alpha\right)\right)\right]>P_{vmin}.\label{e:Constraint}
\end{equation}
where the expectation is taken with respect to the future measurement
sets $Z_{k+1:k+H}\left(\alpha\right)$. 

In general, Monte Carlo integration is used to compute the expected
reward in \eqref{e:Objective} because analytic solutions are not
available. For each control action $\alpha$, this involves drawing
samples $Z_{k+1:k+H}^{\left(i\right)}\left(\alpha\right)$ for $i=1,\dots,N$,
then computing the reward $\mathcal{R}_{k+H}^{\left(i\right)}\left(\alpha\right)$
conditioned on each sample. The samples $Z_{k+1:k+H}^{\left(i\right)}\left(\alpha\right)$
are obtained by first sampling from $\boldsymbol{\pi}_{k}\left(\cdot|Z_{1:k}\right)$,
then propagating each sample through the transition model up to the
horizon time, and finally simulating a set of measurements from time
$k+1$ to time $k+H$ according to the measurement model. An estimate
of the expected reward is given by the mean of the reward over all
the samples,
\begin{equation}
\mathbb{E}\left[\mathcal{R}_{k+H}\left(\alpha\right)\right]\approx\frac{1}{N}\sum_{i=1}^{N}\mathcal{R}_{k+H}^{\left(i\right)}\left(\alpha\right).\label{e:Expected_Reward}
\end{equation}
Since we are sampling directly from the current distribution of $Z_{k+1:k+H}$,
using the true transition and measurement models, this method converges
to the true expectation of the reward as the number of samples is
increased.

In \eqref{e:Expected_Reward}, $\mathcal{R}_{k+H}^{\left(i\right)}\left(\alpha\right)$
is usually computed by Monte Carlo integration, for example \cite{Ristic2010}.
Hence, the variance of the Monte Carlo estimate of the expected reward
will depend on the number of samples used to calculate each $\mathcal{R}_{k+H}^{\left(i\right)}\left(\alpha\right)$,
as well as the number of measurement samples $N$. On the other hand,
a closed form expression for $\mathcal{R}_{k+H}^{\left(i\right)}\left(\alpha\right)$
would lead to a smaller variance in the estimate of the expected reward,
by the principle of Rao-Blackwellization \cite{Casella1996}. 

The constraint \eqref{e:Constraint} is the minimum value of the void
probability up to the control horizon, where the value at time $k+i$
is computed based on the predicted density at that time given measurements
up to time $k$. The constraint is satisfied if this minimum value
is greater than the threshold $P_{vmin}$.

\subsection{Generalised Labelled Multi-Bernoulli Tracking Filter}

This section contains a brief outline of the GLMB Bayes recursion,
which is an essential estimation component in our POMDP-based control
scheme.

Each target is labeled with an ordered pair of integers $\ell=\left(k,i\right)$,
where $k$ is the time of birth, and $i$ is a unique index to distinguish
targets born at the same time. The label space for targets born at
time $k$ is denoted as $\mathbb{L}_{k}$ and the label space for
all targets at time $k$ (including those born prior to $k$) is denoted
as $\mathbb{L}_{0:k}$. Note that $\mathbb{L}_{0:k}=\mathbb{L}_{0:k-1}\cup\mathbb{L}_{k}$.

An existing target at time $k$ has state $\left(x,\ell\right)$ consisting
of the kinematic/feature $x\in\mathcal{\mathbb{X}}$ and label $\ell\in\mathbb{L}_{0:k}$,
i.e. single-target state space $\mathcal{X}$ is the Cartesian product
$\mathcal{\mathbb{X}\times}\mathbb{L}_{0:k}$. An \emph{association
map} at time $k$ is a function $\theta:\mathbb{L}_{0:k}\rightarrow\left\{ 0,1,...,\left|Z\right|\right\} $
such that $\theta\left(\ell\right)=\theta\left(\ell^{\prime}\right)>0$
implies $\ell=\ell^{\prime}$. Such a function can be regarded as
an assignment of labels to measurements, with undetected labels assigned
to $0$. The set of all such association maps is denoted as $\Theta_{k}$,
the subset of association maps with domain $L$ is denoted by $\Theta_{k}\left(L\right)$,
and $\Theta_{0:k}\triangleq\Theta_{0}\times...\times\Theta_{k}$.

In the GLMB filter, the multi-target filtering density at time $k-1$
is a GLMB denoted by 
\begin{equation}
\boldsymbol{\pi}_{k-1}\!\left(\!\boldsymbol{X}|Z_{k-1}\!\right)=\Delta\!\left(\boldsymbol{X}\right)\!\!\!\sum_{c\in\Theta_{0:k-1}}\!\!\!w_{k-1}^{\left(c\right)}\!\left(\mathcal{L}\left(\boldsymbol{X}\right)\right)\!\left[p_{k-1}^{\left(c\right)}\right]^{\boldsymbol{X}}\label{e:GLMB_Previous}
\end{equation}
The set of targets $\boldsymbol{B}_{k}$ born at time $k$ is modelled
by a GLMB with one term: $f_{B,k}\left(\boldsymbol{X}\right)=\Delta\left(\boldsymbol{X}\right)w_{B,k}\left(\mathcal{L}\left(\boldsymbol{X}\right)\right)p_{B,k}^{X}$
(a full GLMB birth can also be easily accommodated) \cite{Vo2013}.
The probability that a target with state $\boldsymbol{x}_{k-1}$ survives
from time $k-1$ to time $k$ is $P_{S,k|k-1}\left(\boldsymbol{x}_{k-1}\right)$.
If a target survives, it transitions to a new state $\boldsymbol{S}_{k}\left(\boldsymbol{x}_{k-1}\right)=\left\{ \left(x_{k},\ell_{k}\right)\right\} $
at time $k$ according to the transition kernel
\begin{align}
 & f_{k|k-1}\left(x_{k},\ell_{k}|x_{k-1},\ell_{k-1}\right)\\
 & \qquad\qquad\qquad=f_{k|k-1}\left(x_{k}|x_{k-1},\ell_{k-1}\right)\delta_{\ell_{k-1}}\left[\ell_{k}\right],\nonumber 
\end{align}
otherwise the target dies and $\boldsymbol{S}_{k}\left(\boldsymbol{x}_{k-1}\right)=\emptyset$.

Under the standard multi-object dynamic model, if the multi-object
filtering density $\boldsymbol{\pi}_{k-1}$ at the previous time is
a GLMB of the form \eqref{e:GLMB_Previous}, then the multi-object
prediction density $\boldsymbol{\pi}_{k|k-1}$ is a GLMB given by
\cite{Vo2013}
\begin{equation}
\boldsymbol{\pi}_{k|k-1}(\boldsymbol{X}|Z_{k-1}\!)=\Delta\!\left(\boldsymbol{X}\right)\!\!\sum_{c\in\Theta_{0:k-1}}\!\!\!w_{k|k-1}^{\left(c\right)}\!\left(\mathcal{L}\left(\boldsymbol{X}\right)\right)\!\left[p_{k|k-1}^{\left(c\right)}\right]^{\boldsymbol{X}},\label{e:GLMB_Prediction}
\end{equation}
where
\begin{align*}
w_{k|k-1}^{\left(c\right)}\left(L\right) & =w_{S,k|k-1}^{\left(c\right)}\left(L\cap\mathbb{L}_{0:k-1}\right)w_{B,k}\left(L\cap\mathbb{L}_{k}\right),\\
p_{k|k-1}^{\left(c\right)}\left(x,\ell\right) & =1_{\mathbb{L}_{0:k-1}}\!\!\left(\ell\right)p_{S,k|k-1}^{\left(c\right)}\!\left(x,\ell\right)\!+\!1_{\mathbb{L}_{k}}\!\left(\ell\right)p_{B,k}\!\left(x,\ell\right),\\
w_{S,k|k-1}^{\left(c\right)}\left(L\right) & =\left[\bar{P}_{S,k|k-1}^{(c)}\right]^{L}\!\sum_{I\supseteq L}\!\left[1\!-\!\bar{P}_{S,k|k-1}^{\left(c\right)}\right]^{I-L}\!\!w_{k-1}^{\left(c\right)}\!\left(I\right),\\
\bar{P}_{S,k|k-1}^{\left(c\right)}\left(\ell\right) & =\left\langle P_{S,k|k-1}\left(\cdot,\ell\right),p_{k-1}^{\left(c\right)}\left(\cdot,\ell\right)\right\rangle ,\\
p_{S,k|k-1}^{\left(c\right)}\left(x,\ell\right) & =\frac{\left\langle P_{S,k|k-1}\!\left(\cdot,\ell\right)f_{k|k-1}\!\left(x,\ell|\cdot,\ell\right),p_{k-1}^{\left(c\right)}\!\left(\cdot,\ell\right)\right\rangle }{\bar{P}_{S,k|k-1}^{\left(c\right)}\left(\ell\right)},
\end{align*}

Each target is detected with probability $P_{D,k}\left(\boldsymbol{x}_{k}\right)$,
and if detected generates a singleton measurement $D_{k}\left(\boldsymbol{x}_{k}\right)=\left\{ z_{k}\right\} $
with probability density $g_{k}\left(z_{k}|\boldsymbol{x}_{k}\right)$,
otherwise it generates the empty set $D_{k}=\emptyset$. The RFS of
false alarms $F_{k}$ is Poisson with intensity function $\kappa\left(\cdot\right)$.
Under the standard multi-object observation model, if the predicted
multi-object density is a GLMB of the form \eqref{e:GLMB_Prediction},
the posterior multi-object density $\boldsymbol{\pi}_{k}\left(\cdot|Z_{k}\right)$
is a GLMB given by
\begin{align}
 & \boldsymbol{\pi}_{k}\left(\boldsymbol{X}\mathbf{|}Z_{k}\right)=\label{e:GLMB_Update}\\
 & \qquad\Delta\!\left(\boldsymbol{X}\right)\!\!\sum_{c\in\Theta_{0:k-1}}\sum\limits _{\theta\in\Theta_{k}}\!w_{k}^{\left(c,\theta\right)}\!\!\left(\mathcal{L}\left(\boldsymbol{X}\right)|Z_{k}\right)\!\left[p^{\left(c,\theta\right)}\left(\cdot|Z_{k}\right)\right]^{\boldsymbol{X}},\nonumber 
\end{align}
where
\begin{align*}
w_{k}^{\left(c,\theta\right)}\left(L|Z\right) & \propto1_{\Theta_{k}\left(L\right)}\left(\theta\right)\left[\bar{\Psi}_{Z,k}^{\left(c,\theta\right)}\right]^{L}w_{k|k-1}^{\left(c\right)}\left(L\right),\\
\bar{\Psi}_{Z,k}^{\left(c,\theta\right)}\left(\ell\right) & =\left\langle \Psi_{Z,k}^{\left(\theta\right)}\left(\cdot,\ell\right),p_{k|k-1}^{\left(c\right)}\left(\cdot,\ell\right)\right\rangle ,\\
\Psi_{Z,k}^{\left(\theta\right)}\left(x,\ell\right) & =\left[1-P_{D,k}\left(x,\ell\right)\right]^{\delta\left[\theta\left(\ell\right)\right]}\\
 & \qquad\times\left[\frac{P_{D,k}\left(x,\ell\right)g_{k}\left(z_{\theta\left(\ell\right)}|x,\ell\right)}{\kappa\left(z_{\theta\left(\ell\right)}\right)}\right]^{1-\delta\left[\theta\left(\ell\right)\right]}\\
p_{k}^{\left(c,\theta\right)}\left(x,\ell|Z\right) & =\frac{\Psi_{Z,k}^{\left(\theta\right)}\left(x,\ell\right)p_{k|k-1}^{\left(c\right)}\left(x,\ell\right)}{\bar{\Psi}_{Z,k}^{\left(c,\theta\right)}\left(\ell\right)}.
\end{align*}
Note that in this application, these functions will all depend on
the control action $\alpha$, which has been omitted for compactness
of notation.

The GLMB density is thus a conjugate prior with respect to the standard
multi-object likelihood function and is also closed under the multi-object
prediction. Consequently, starting with an initial prior density in
GLMB form, under the standard data and dynamic model, the posterior
density at any time is also a GLMB. The recursion above is the first
exact closed form solution to the Bayes multi-target filter. In \cite{Vo2014}
an implementation of the GLMB filter based on discarding `insignificant'
components was detailed, and it was shown that such truncation minimizes
the $L_{1}$ error in the multi-target density. This algorithm has
a worst case complexity that is cubic in the number of observations.

\subsection{Reward Function\label{s:Reward_Function}}

For the reason discussed in Section \ref{s:Control_Strategy}, the
existence of a closed form reward function is desirable in POMDPs.
This would be particularly beneficial in this application, since the
difference between the expected rewards of the various control actions
can be quite small, and may become obscured by the variance induced
by the Monte Carlo estimation \eqref{e:Expected_Reward}. Any reduction
in this variance will clearly help in correctly identifying the optimal
control action.

For the case of the GLMB, common information divergence measures such
as the Kullback-Liebler or Rényi divergences cannot be expressed in
analytical form. Thus, their use in this problem would require Monte
Carlo integration, resulting in a higher variance in the expected
reward, as well as increased computational load. To alleviate this,
we use the Cauchy-Schwarz divergence between prior and posterior GLMB
densities as the reward function, i.e.
\begin{align}
 & \mathcal{R}_{k+H}^{\left(i\right)}\left(\alpha\right)=D_{CS}\Big(\boldsymbol{\pi}_{k+H}\left(\boldsymbol{X}|Z_{1:k}\right),\label{e:Sample_Reward}\\
 & \qquad\qquad\qquad\qquad\quad\boldsymbol{\pi}_{k+H}\Big(\boldsymbol{X}|Z_{1:k},Z_{k+1:k+H}^{\left(i\right)}\left(\alpha\right)\Big)\Big).\nonumber 
\end{align}

\subsection{Constraint\label{s:Constraint}}

To enforce the constraint, we compute the void probability over an
exclusion region around the sensor, for the predicted GLMB density
at each time step up to the horizon. We use a circular exclusion region,
centered at the sensor location, with radius $r_{V}$. Evaluation
of the void probability requires integrating each single-object density
in the GLMB over the exclusion region. For a 2-dimensional Gaussian
pdf and a circular region, this does not have an analytical solution.
Hence, we use adaptive cubature to approximate these integrals, before
using them to compute the void probability.

The goal is to ensure that the separation between sensor and targets
always exceeds $r_{V}$, i.e. the control action $\alpha$ is feasible
if the constraint \eqref{e:Constraint} is satisfied. That is, for
each action we find the minimum void probability up to the horizon,
and enforce the constraint that this minimum must exceed the threshold
$P_{vmin}$, otherwise the action cannot be selected.

\subsection{Simulation Results\label{s:Simulation_Results}}

In this section, the control strategy is applied to the problem of
observer trajectory optimization for multi-target tracking. This application
involves a single sensor that provides bearing and range measurements,
where the noise on the measured bearings is constant for all targets,
but the range noise is state-dependent, increasing as the true range
between the sensor and target increases. The detection probability
is also range-dependent, reducing as the range increases. Targets
closer to the sensor are therefore detected with both higher probability
and accuracy, and vice versa for targets that are further away.

For this problem, the state-dependency of the measurement noise and
detection probability will be the main effect driving the control,
and one would expect the algorithm to move the sensor towards the
targets, in order to minimise noise and maximise the detection probability.
However, in the presence of multiple targets, this can easily lead
to conflicting control influences. The goal of the control algorithm
is to resolve these conflicts, by attempting to provide a decision
that optimizes the multi-target estimation performance in an overall
sense.

The target kinematics are modelled using 2D Cartesian position and
velocity vectors $x_{k}=\left[\begin{array}{cccc}
t_{x,k} & \dot{t}_{x,k} & t_{y,k} & \dot{t}_{y,k}\end{array}\right]^{T}$, and they are assumed to move according to the following discrete
white noise acceleration model,
\begin{align}
x_{k} & =Fx_{k-1}+\Gamma v_{k-1},\label{e:Dynamic_Model}\\
F & =\left[\begin{array}{cc}
1 & T\\
0 & 1
\end{array}\right]\otimes I_{2},\qquad\Gamma=\left[\begin{array}{c}
T^{2}/2\\
T
\end{array}\right]\otimes I_{2}\nonumber 
\end{align}
where $T$ is the sampling period, $v_{k-1}\sim\mathcal{N}\left(0,Q\right)$
is a $2\times1$ independent and identically distributed Gaussian
process noise vector with $Q=\sigma_{v}^{2}I_{2}$, where $\sigma_{v}$
is the standard deviation of the target acceleration. The sensor measures
the target bearing and range, where the measurement corresponding
to a target state $x_{k}$ and sensor position $u_{k}=\left[\begin{array}{cc}
s_{x,k} & s_{y,k}\end{array}\right]$ at time $k$ is given by 
\begin{align}
z_{k} & =h\left(x_{k},u_{k}\right)+w_{k}\left(x_{k},u_{k}\right)\label{e:Measurement_Model}
\end{align}
where $w_{k}\left(x_{k},u_{k}\right)\sim\mathcal{N}\left(0,\text{diag}\left(\left[\begin{array}{cc}
\sigma_{\theta}^{2} & \sigma_{r}^{2}\left(x_{k},u_{k}\right)\end{array}\right]\right)\right)$ is a $2\times1$ Gaussian measurement noise vector, and the measurement
function $h$ is given by 
\begin{equation}
h\left(x_{k},u_{k}\right)=\left[\!\begin{array}{c}
\arctan\left(\frac{t_{y,k}-s_{y,k}}{t_{x,k}-s_{x,k}}\right)\\
\sqrt{\left(t_{x,k}\!-\!s_{x,k}\right)^{2}\!+\!\left(t_{y,k}\!-\!s_{y,k}\right)^{2}}
\end{array}\!\right].\label{e:Measurement_Equation}
\end{equation}

The variance of the bearing measurement noise $\sigma_{\theta}^{2}$
is a fixed constant for all targets, but the variance of the range
noise $\sigma_{r}^{2}$ is a function of the target and sensor states.
In this example, we model the range noise in a piecewise manner as
follows, in which $\mathcal{D}\left(x_{k},u_{k}\right)$ denotes the
true distance between a target with state $x_{k}$ and the sensor
with state $u_{k}$,
\begin{equation}
\sigma_{r}^{2}\left(\!x_{k},u_{k}\!\right)\!=\!\begin{cases}
\!\left(\eta R_{1}\right)^{2}, & \mathcal{D}\left(\!x_{k},u_{k}\!\right)\leq R_{1}\\
\!\left(\!\eta\mathcal{D}\left(\!x_{k},u_{k}\!\right)\!\right)^{2}\!, & R_{1}\!<\!\mathcal{D}\left(\!x_{k},u_{k}\!\right)\!<\!R_{2}\\
\!\left(\eta R_{2}\right)^{2}, & \mathcal{D}\left(\!x_{k},u_{k}\!\right)\geq R_{2}
\end{cases},\label{e:Range_Noise_Profile}
\end{equation}
i.e. the noise standard deviation is the true range mutiplied by the
factor $\eta$, but the minimum is capped at $\eta R_{1}$, and the
maximum is capped at $\eta R_{2}$. The detection probability is modeled
using the following function of the true range,
\begin{equation}
p_{D}\left(x_{k},u_{k}\right)=\frac{\mathcal{N}\left(\mathcal{D}\left(x_{k},u_{k}\right);0,\sigma_{D}^{2}\right)}{\mathcal{N}\left(0;0,\sigma_{D}^{2}\right)},\label{e:Detect_Prob_Profile}
\end{equation}
where $\sigma_{D}$ controls the rate at which the detection probability
drops off as the range increases.

To illustrate the performance of the control, we apply it to two different
simulated scenarios. The first has a time-varying number of targets,
and demonstrates how the algorithm adapts to the changing conditions
over time. The second scenario consists of targets which are scattered
in several different locations and moving in different directions.
In scenario 1, the expected control behaviour is fairly clear from
looking at the target-observer geometry. However, in scenario 2, the
expected behaviour is not so obvious.

For both scenarios, the sensor sampling interval is $T=10\text{s}$,
the clutter rate is $100$ per scan, the detection probability spread
parameter is $\sigma_{D}=20\text{km}$, the process noise on the target
trajectories is $\sigma_{v}=0.01\text{m/s}^{2}$, the bearing measurement
noise is $\sigma_{\theta}=2^{\circ}$, and the parameters of the range
measurement noise are $R_{1}=1\text{km}$, $R_{2}=10\text{km}$ and
$\eta=0.1$. The space of possible control actions is discretized
at $20^{\circ}$ intervals, i.e. the alllowed course changes are $\left\{ -180^{\circ},-160^{\circ},\dots,0^{\circ},\dots,160^{\circ},180^{\circ}\right\} $.
For the control calculations, the number of samples is $N=50$, the
sensor sampling interval is $T=80s$, and the horizon length is $H=5$
(i.e. the effective control lookahead is 400s). The exclusion radius
for the void probability calculation is $r_{v}=1\text{km}$, and the
void probability threshold is $P_{vmin}=0.95$.

\subsubsection{Scenario 1}

\begin{figure}[H]
\begin{centering}
\includegraphics[width=0.95\columnwidth]{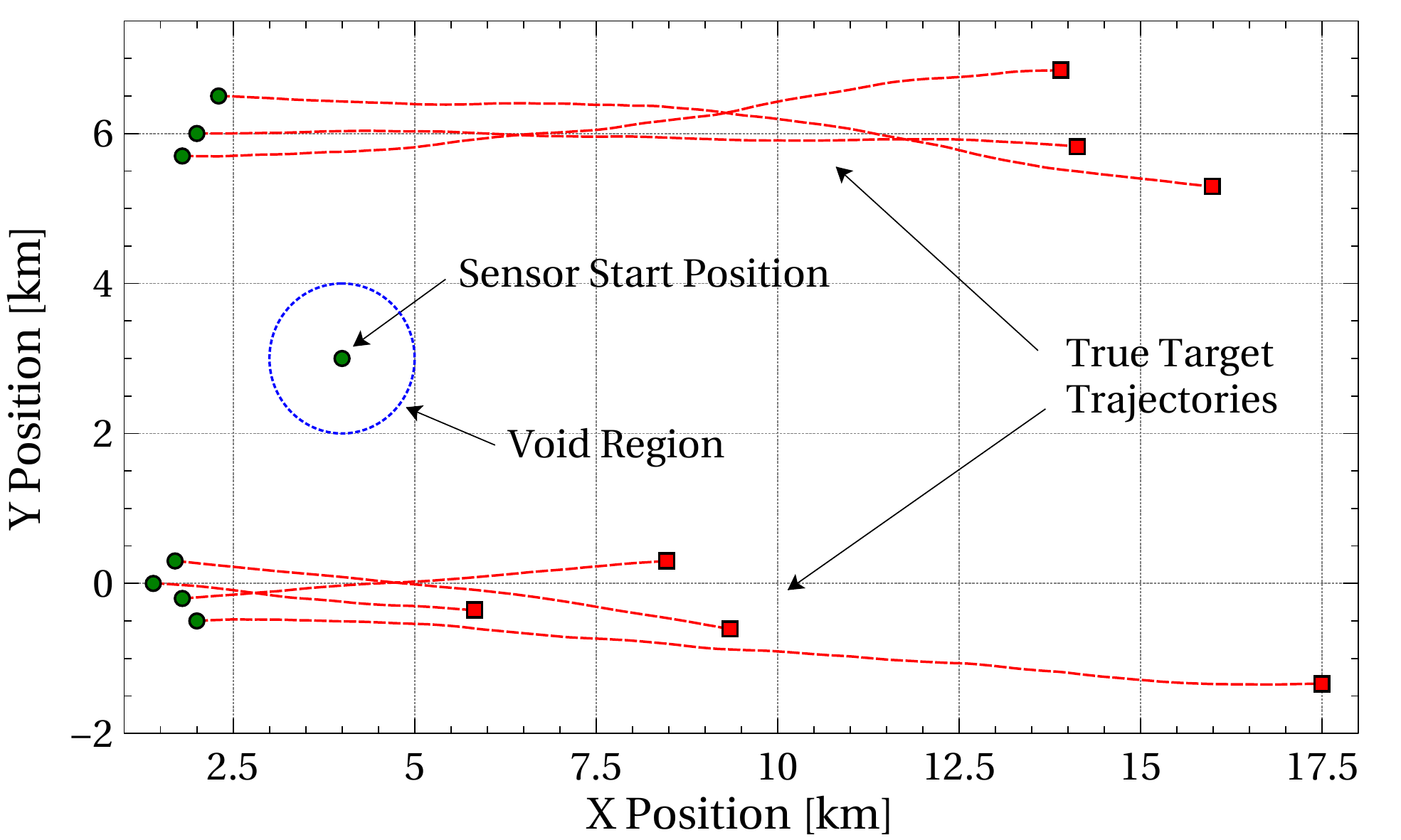}
\par\end{centering}

\caption{True targets trajectories in scenario 1. Note that the targets appear
and disappear at different times, which is not represented on the
plot.}
\label{f:Scenario_1_Geometry}
\end{figure}

The first scenario runs for 4000 seconds and consists of seven targets,
six of which enter the scene during the first 250s, with one more
appearing at time 1700s. Three of the targets terminate between times
1300s and 1600s. The sensor platform is stationary for the first 400s,
then starts moving with constant speed of 7m/s, undergoing course
changes every 400s in order to improve the target estimates. The true
target trajectories are depicted in Figure \ref{f:Scenario_1_Geometry}.

\begin{figure}[H]
\begin{centering}
\includegraphics[width=1\columnwidth]{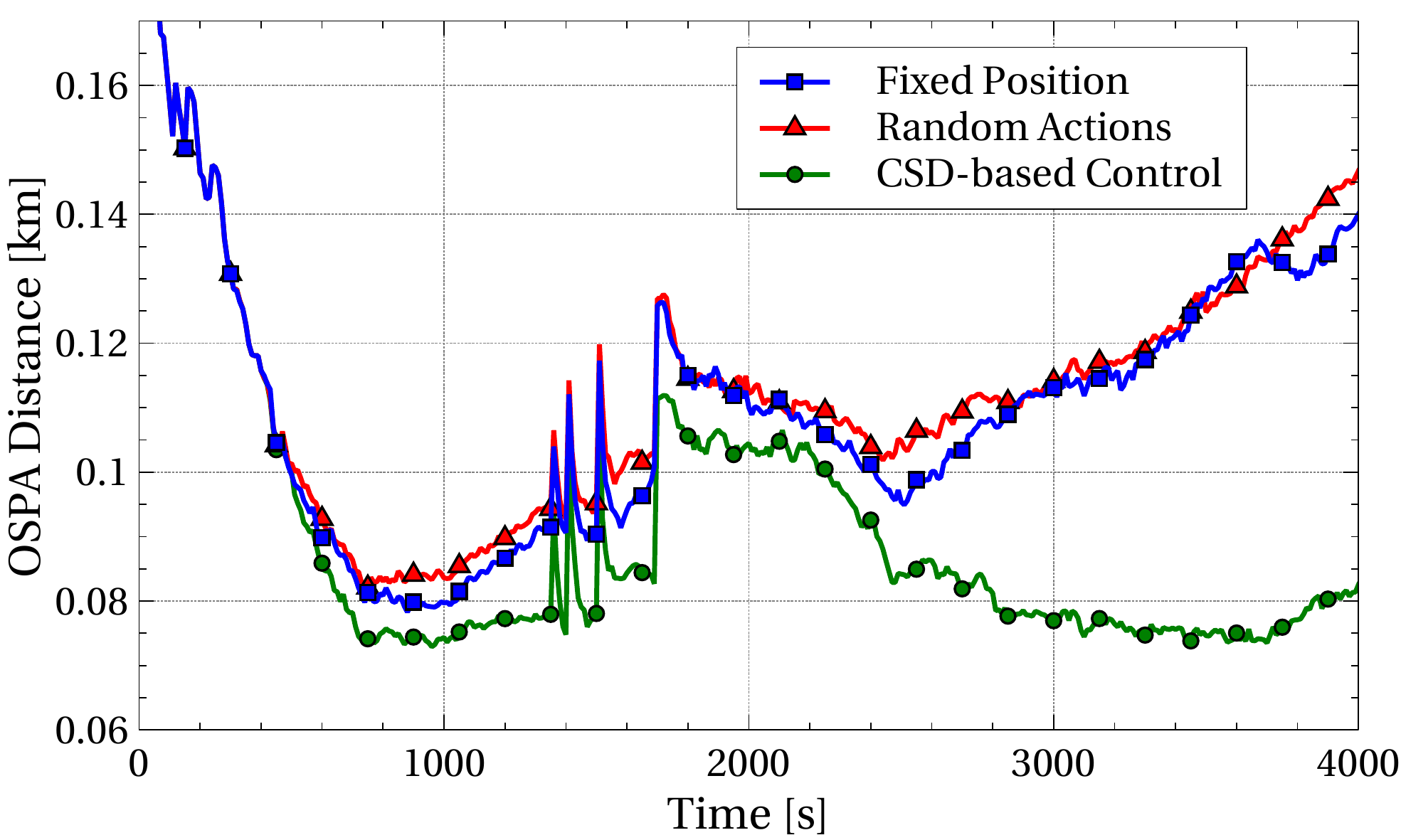}
\par\end{centering}

\caption{Comparison of OSPA versus time for the cases of fixed sensor location,
randomised control actions, and control based on the Cauchy-Schwarz
divergence / void probability.}
\label{f:Scenario_1_OSPA}
\end{figure}

To evaluate the control performance, we compare the estimation accuracy
under the proposed control scheme, against the case of a stationary
sensor, and the case where the sensor performs randomly chosen course
changes. For each of the three cases, we have performed 100 Monte
Carlo runs, and used a multi-target miss distance known as the optimal
sub-pattern assignment (OSPA) metric \cite{Schumacher2008}, to quantify
the positional error between the filter estimates and the ground truth.
Figure \ref{f:Scenario_1_OSPA} shows the average of the OSPA distance
versus time where the OSPA cutoff parameter is $c=200\text{m}$ and
the order parameter is $p=2$. These results show that the proposed
control strategy provides the best estimation performance. In the
cases where the sensor is stationary or undergoing randomly chosen
actions, the performance is significantly worse, since they have no
mechanism for positioning the sensor in the most favourable location.
This demonstrates that the use of Cauchy-Schwarz divergence as the
reward function has been effective at reducing the estimation error
of the system. 

\begin{figure}[H]
\begin{centering}
\includegraphics[width=1\columnwidth]{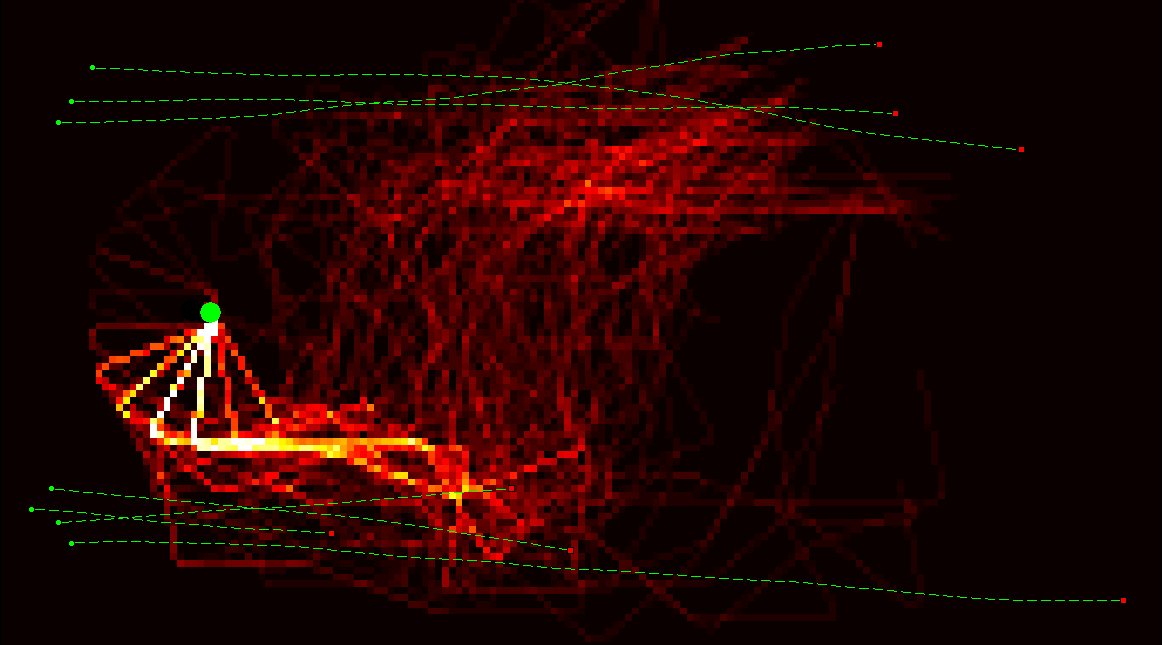}
\par\end{centering}

\caption{This heatmap shows the sensor location over 100 Monte Carlo runs.
Brighter colors represent locations that were more frequently visited.
The sensor usually starts by moving towards the four targets at the
bottom. After some of those targets become terminated, the sensor
moves towards the three targets at the top. A few exceptions to this
behaviour can be seen, but the general trend is clearly visible.}

\label{f:Scenario_1_Heatmap}
\end{figure}

Figure \ref{f:Scenario_1_Heatmap} shows a heatmap summarizing the
paths taken by the sensor over the 100 Monte Carlo runs. From this
diagram, the general trend of the controlled sensor's trajectory can
be observed. Intuition would suggest that the sensor should move closer
to the areas with the higher concentration of targets, which agrees
with the trend shown in Figure \ref{f:Scenario_1_Heatmap}.

\begin{figure}[H]
\noindent \begin{centering}
\begin{minipage}[c][1\totalheight][t]{0.4\columnwidth}%
\begin{center}
\includegraphics[width=1\columnwidth]{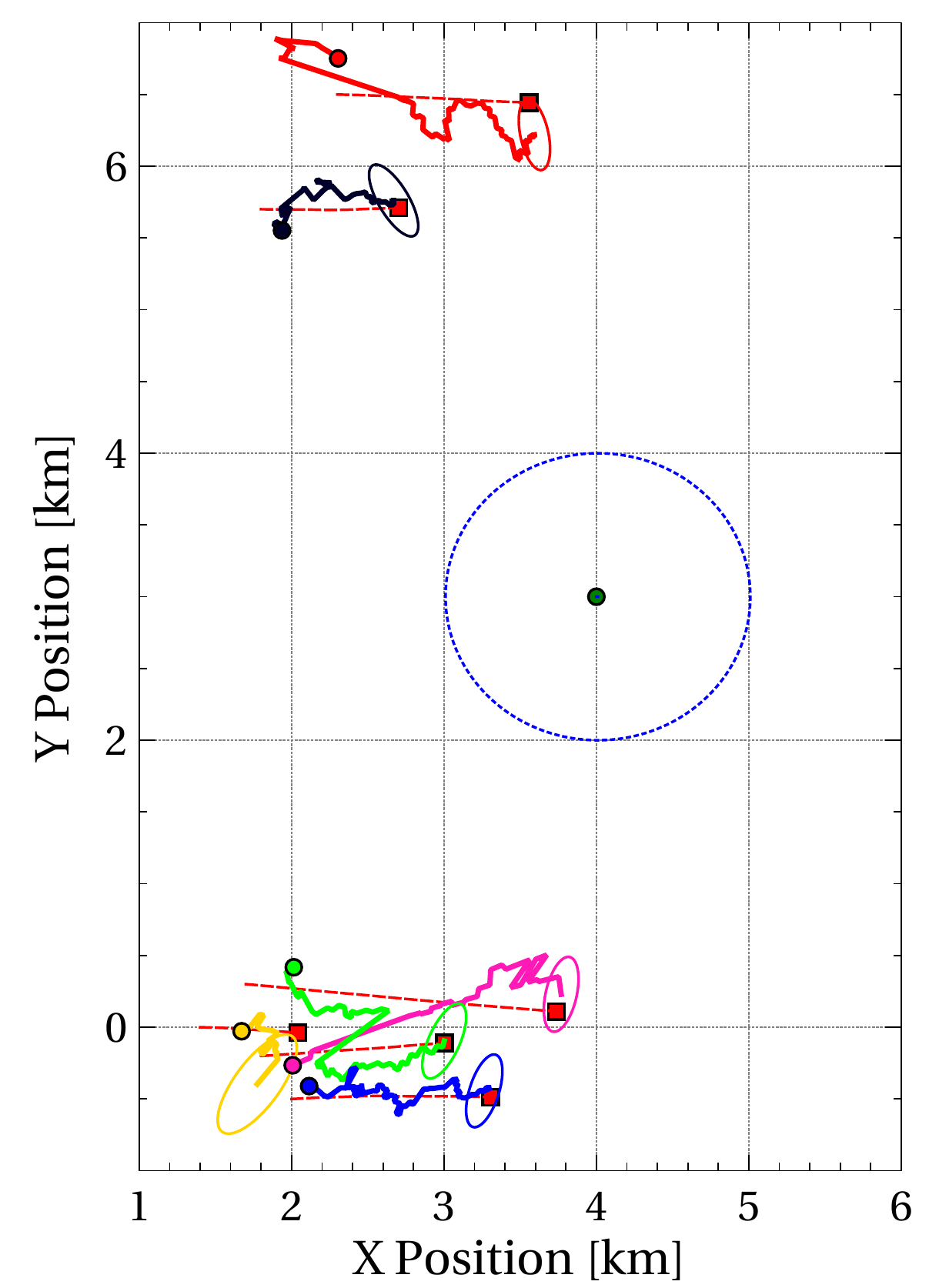}
\par\end{center}%
\end{minipage}%
\begin{minipage}[c][1\totalheight][t]{0.6\columnwidth}%
\begin{center}
\includegraphics[width=1\columnwidth]{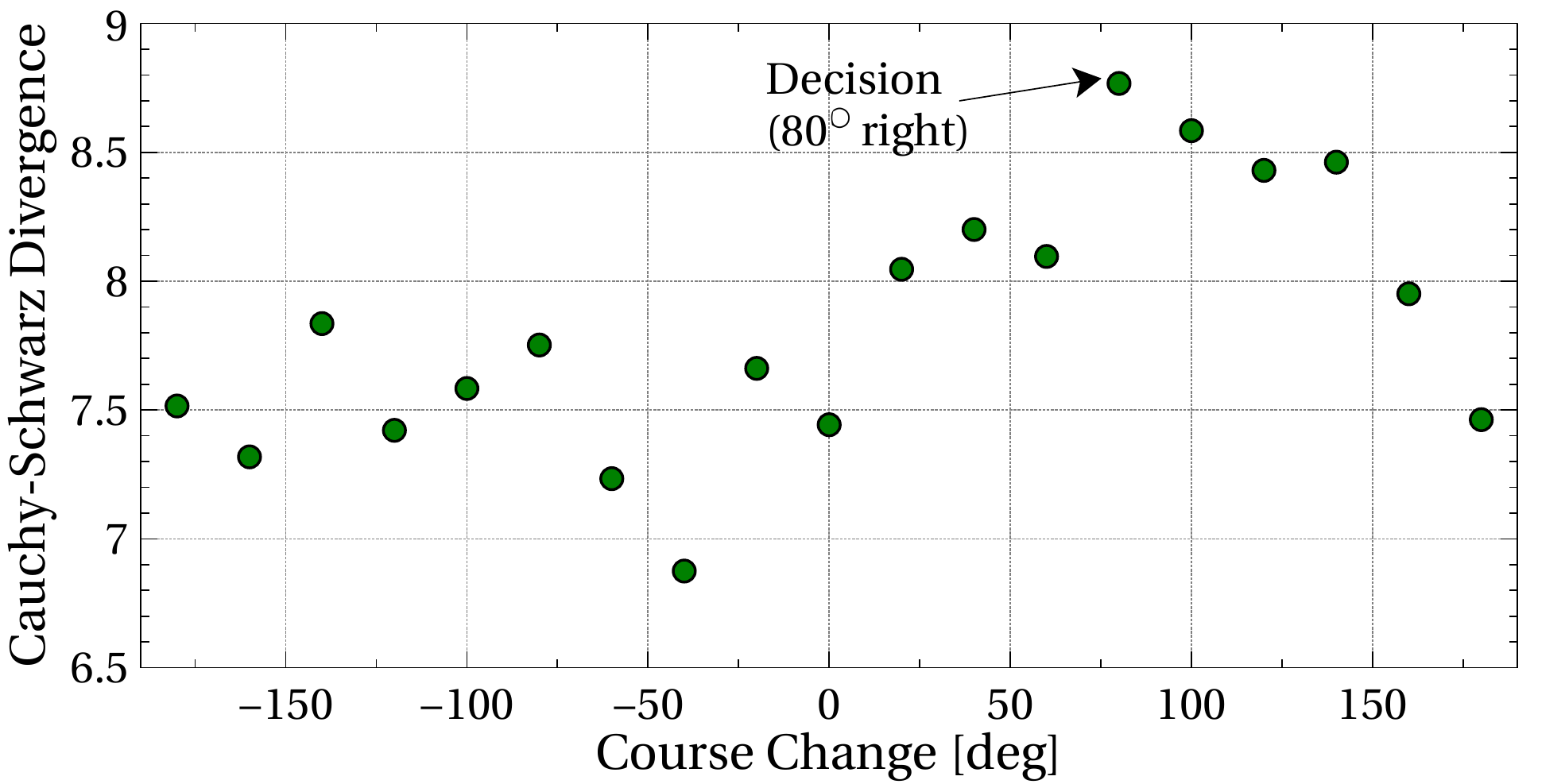}
\par\end{center}%
\end{minipage}
\par\end{centering}

\caption{Scenario geometry and reward curve at the time of the first decision
(400s). The sensor platform is stationary for the first 400s, and
pointing towards the right. The first decision made by the control
algorithm is to turn $80^{\circ}$ to the right, which takes the sensor
towards the group of four targets.}

\label{f:Scenario_1_Decision_1}
\end{figure}

\begin{figure}[H]
\begin{centering}
\includegraphics[width=0.85\columnwidth]{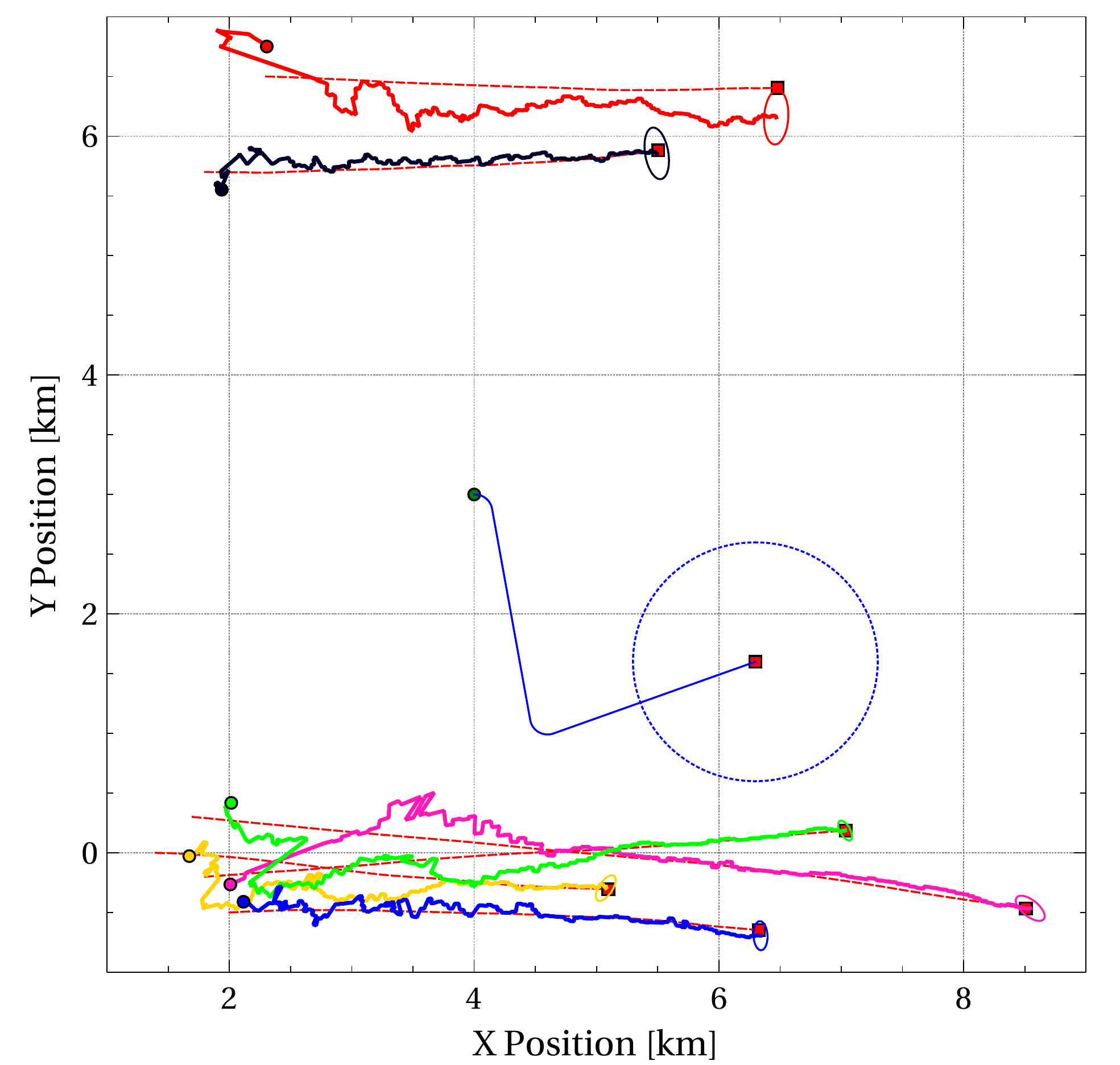}
\par\end{centering}

\begin{centering}
\includegraphics[width=0.85\columnwidth]{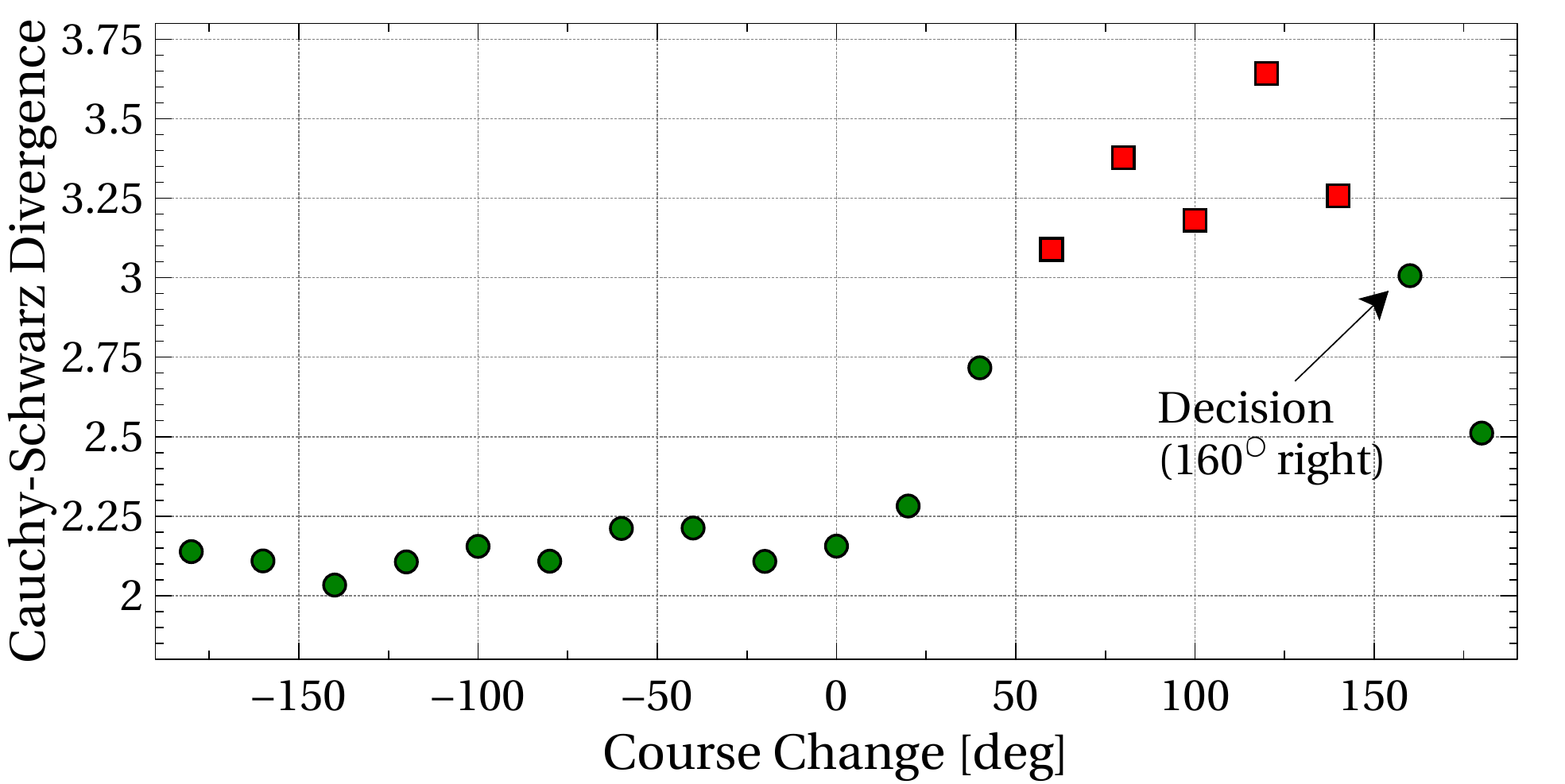}
\par\end{centering}

\begin{centering}
\includegraphics[width=0.85\columnwidth]{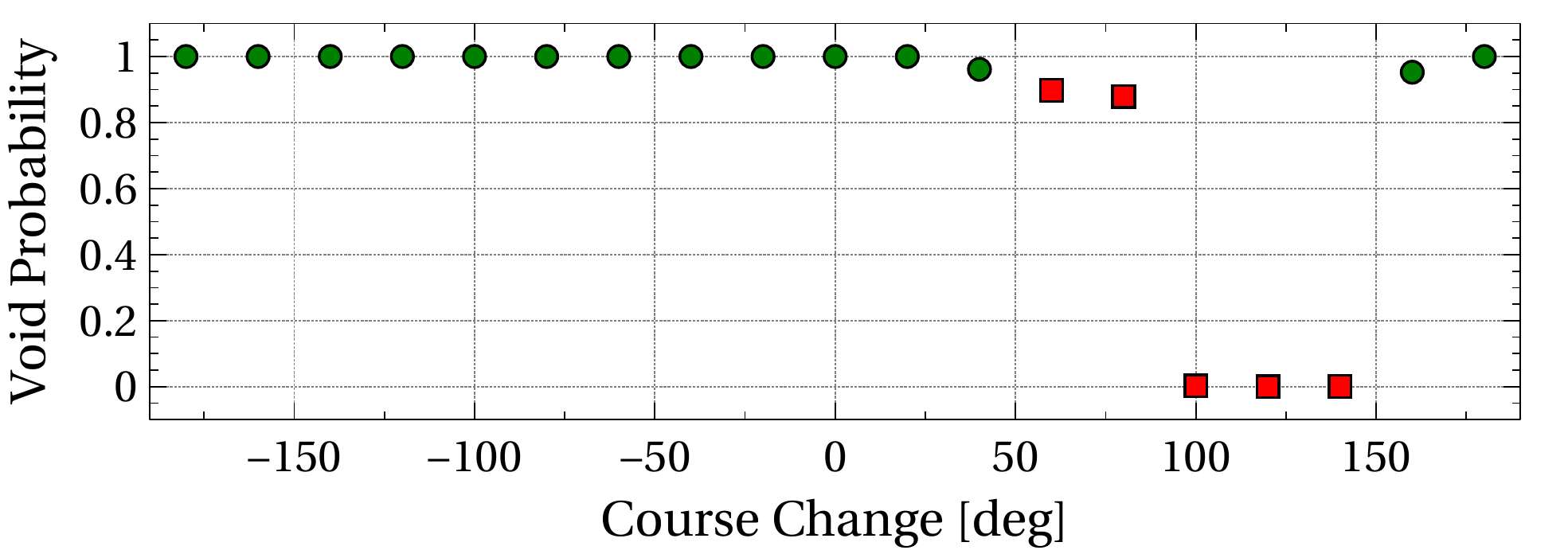}
\par\end{centering}

\caption{Scenario geometry, reward curve and void probability curve at the
time of the third decision (1200s). Four of the manoeuvres do not
satisfy the constraint, because they would result in high probability
of targets getting too close to the sensor. Excluding these, the best
remaining decision is to turn $160^{\circ}$ to the right.}

\label{f:Scenario_1_Decision_3}
\end{figure}

\begin{figure}[H]
\begin{centering}
\includegraphics[width=0.9\columnwidth]{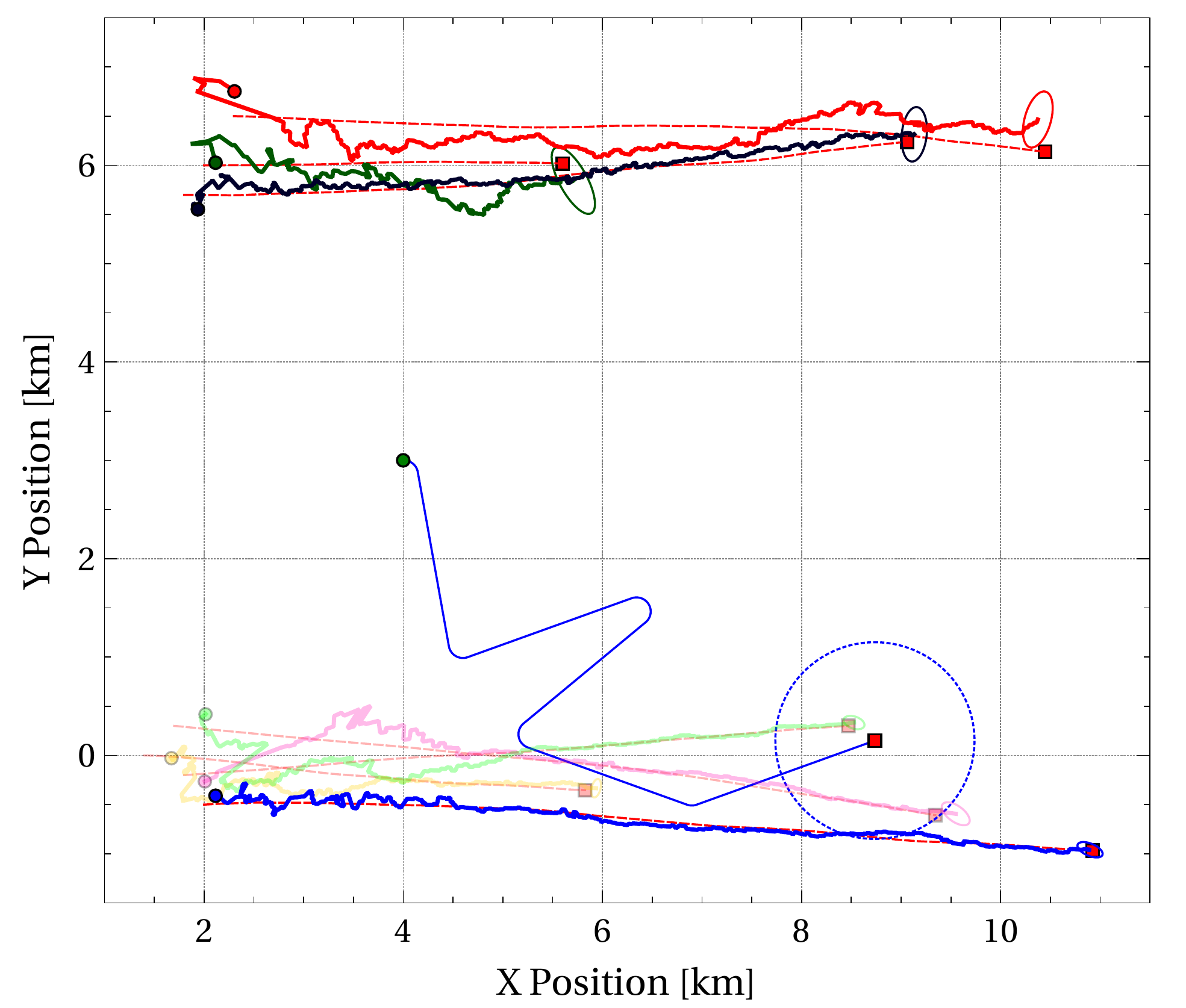}
\par\end{centering}

\begin{centering}
\includegraphics[width=0.9\columnwidth]{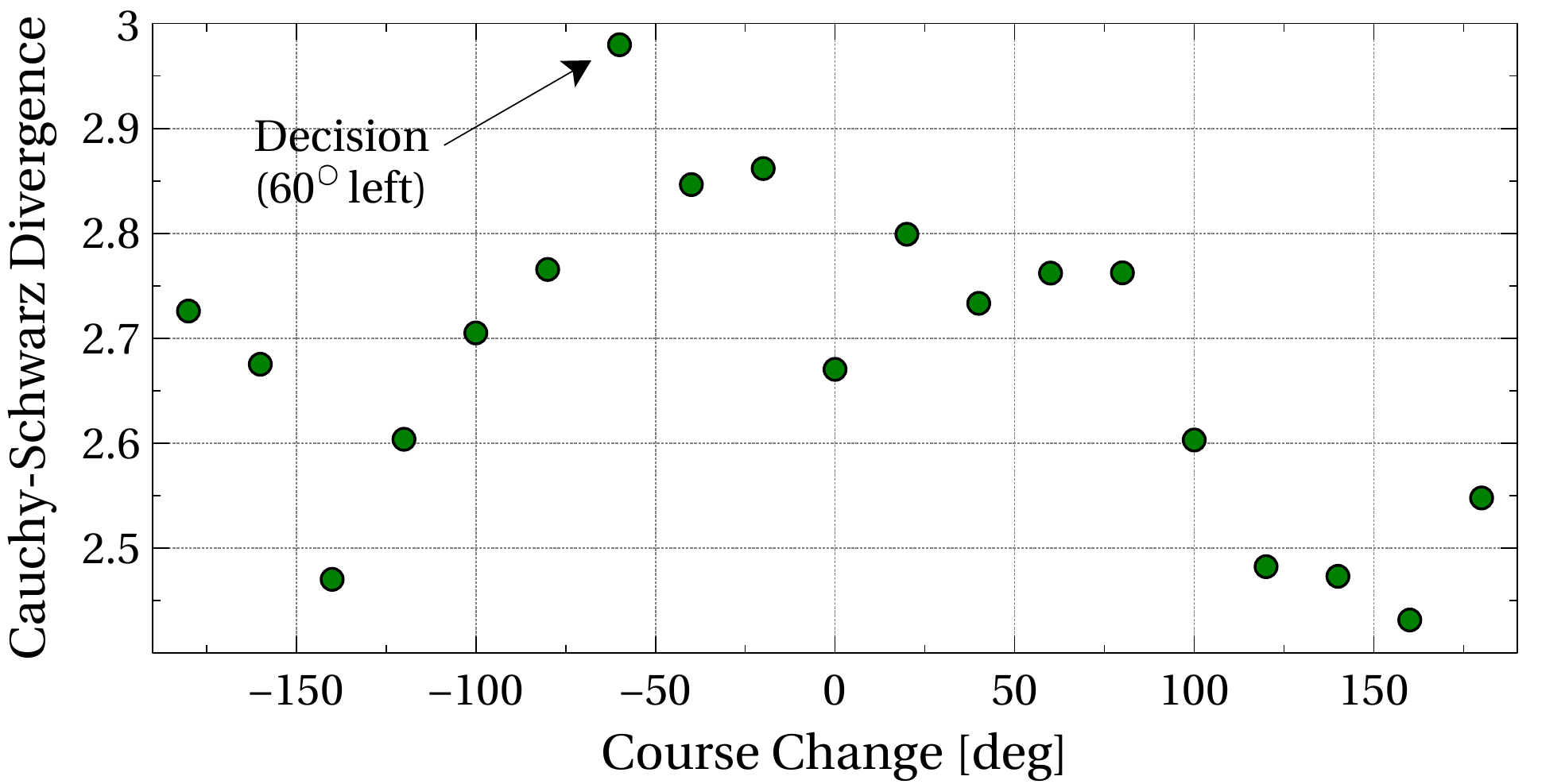}
\par\end{centering}

\caption{Scenario geometry and reward curve at the time of the sixth decision
(2400s). Three of the targets at the bottom have terminated since
time 1200s (as indicated by the faint tracks), and one additional
target has appeared at the top. As a result, the algorithm decides
to turn $60^{\circ}$ left, taking it closer to the three targets
at the top.}

\label{f:Scenario_1_Decision_6}
\end{figure}

\begin{figure}[H]
\begin{centering}
\includegraphics[width=0.9\columnwidth]{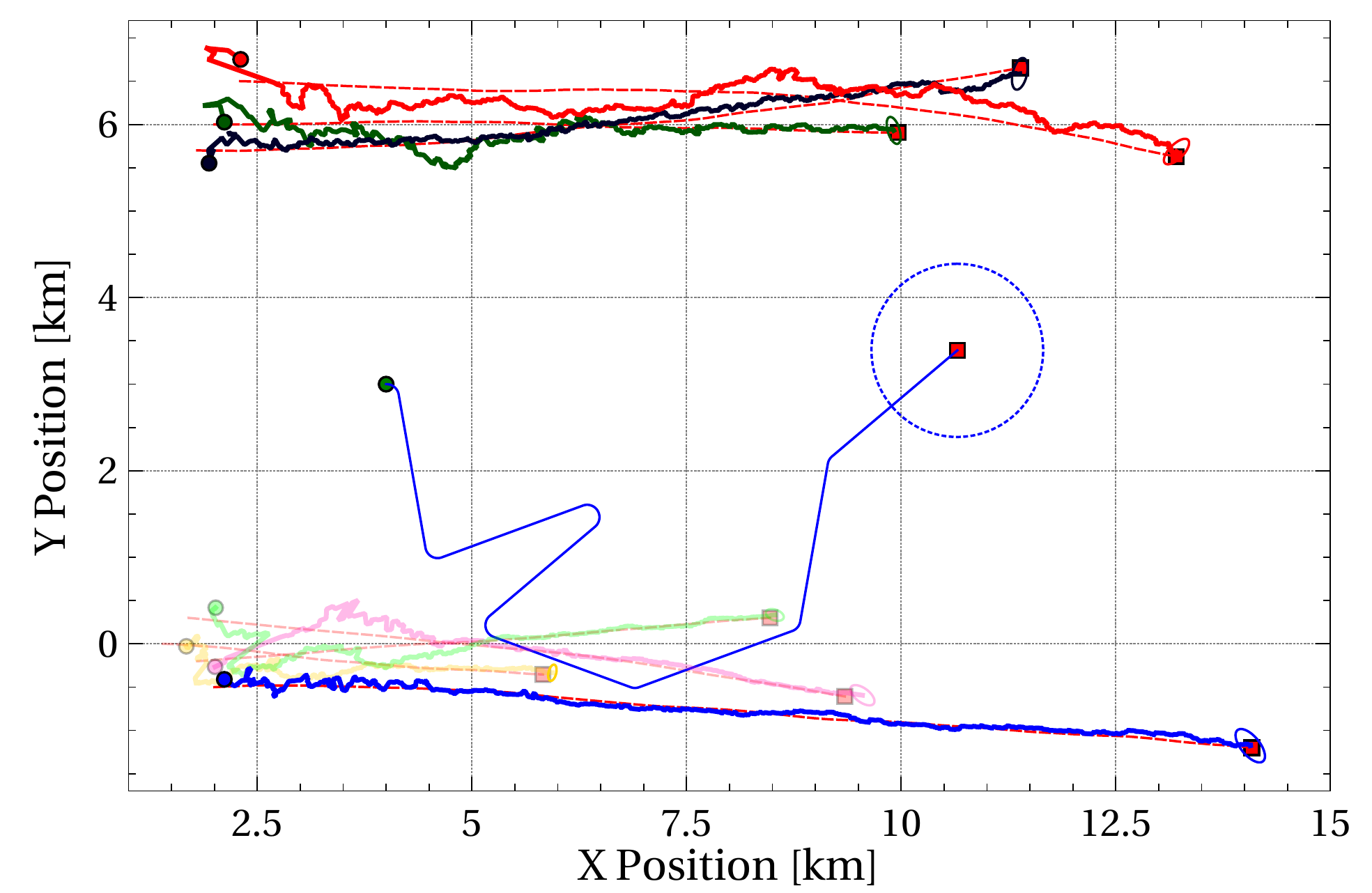}
\par\end{centering}

\begin{centering}
\includegraphics[width=0.9\columnwidth]{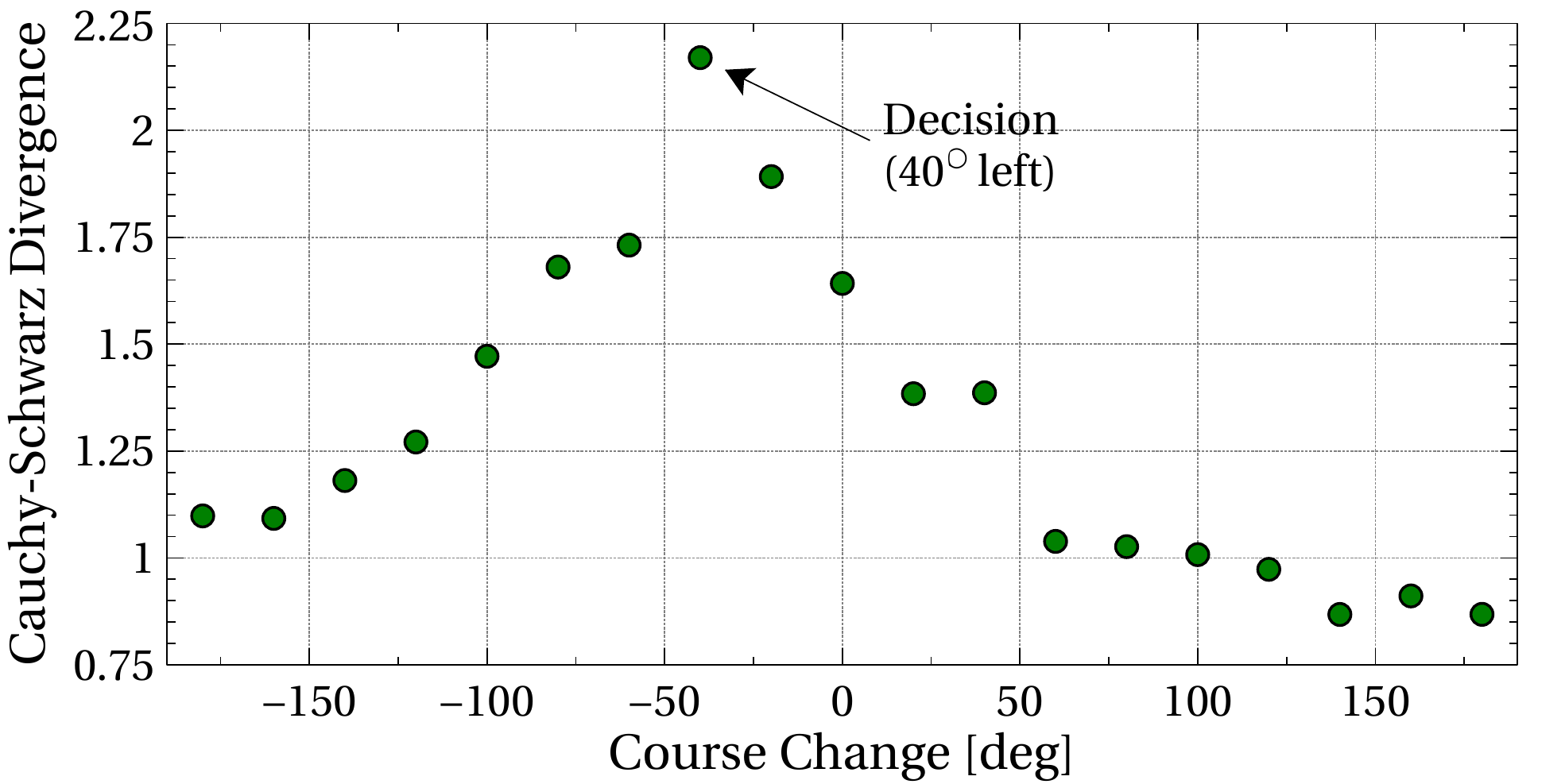}
\par\end{centering}

\caption{Scenario geometry and reward curve at the time of the eigth decision
(3200s). The algorithm decides on a slight turn to the left, which
takes it closer to the group of three targets.}

\label{f:Scenario_1_Decision_8}
\end{figure}

\begin{figure}[H]
\begin{centering}
\includegraphics[width=1\columnwidth]{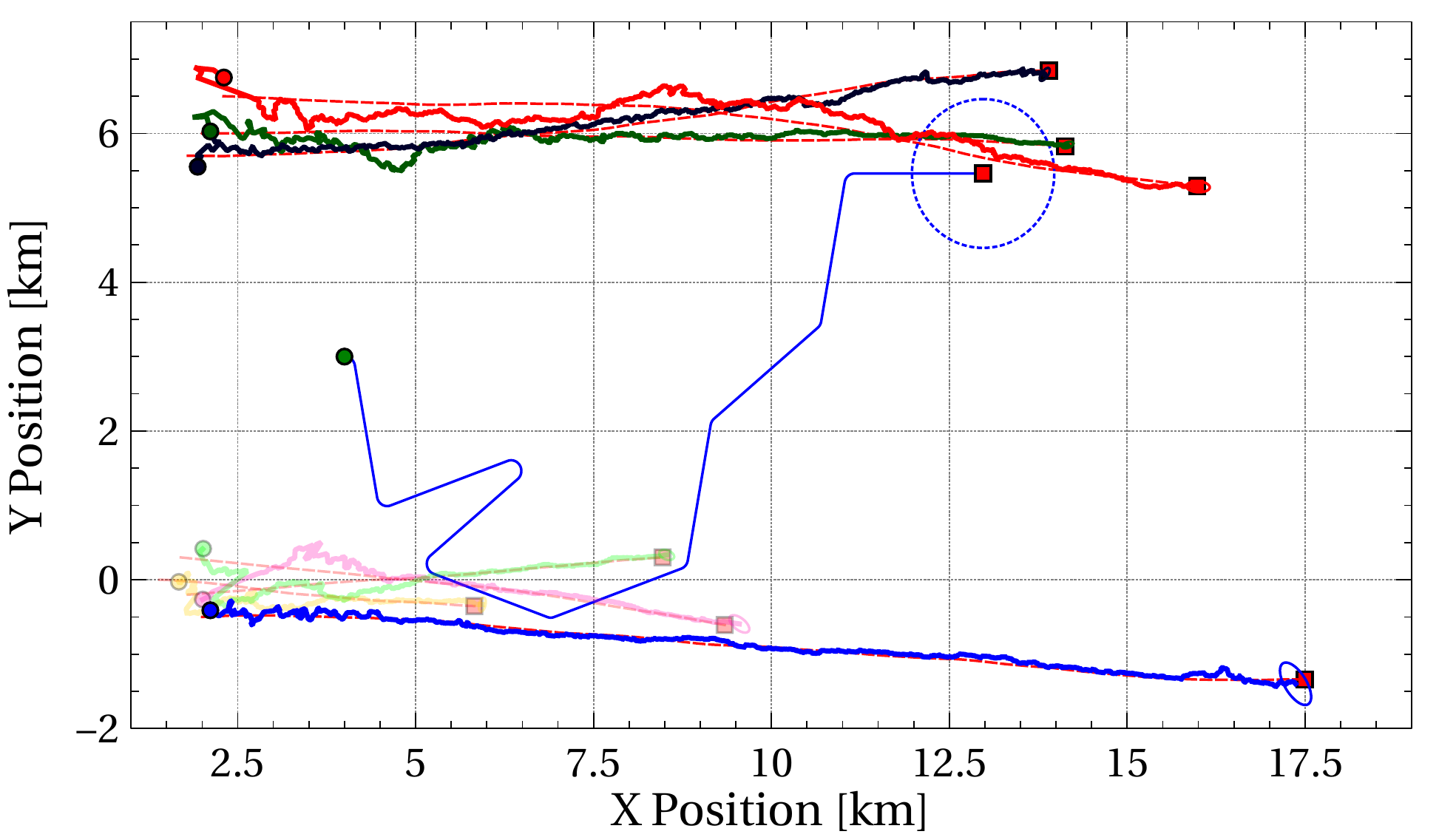}
\par\end{centering}

\caption{Scenario at time 4000s. The sensor is now following the group of three
targets at the top.}

\label{f:Scenario_1_Final}
\end{figure}

To demonstrate the operation of the control scheme, we now show a
single run which exhibits the typical control behaviour. We have shown
the scenario geometry and expected Cauchy-Schwarz divergence for each
potential action at five different time instants; at 400s when the
first decision is made (Figure \eqref{f:Scenario_1_Decision_1}),
the third decision at time 1200s (Figure \ref{f:Scenario_1_Decision_3}),
the sixth decision at time 2400s (Figure \ref{f:Scenario_1_Decision_6}),
the eigth decision at time 3200s (Figure \ref{f:Scenario_1_Decision_8}),
and finally, at the end of the scenario at time 4000s (Figure \ref{f:Scenario_1_Final}). 

\subsubsection{Scenario 2}

\begin{figure}[H]
\begin{centering}
\includegraphics[width=1\columnwidth]{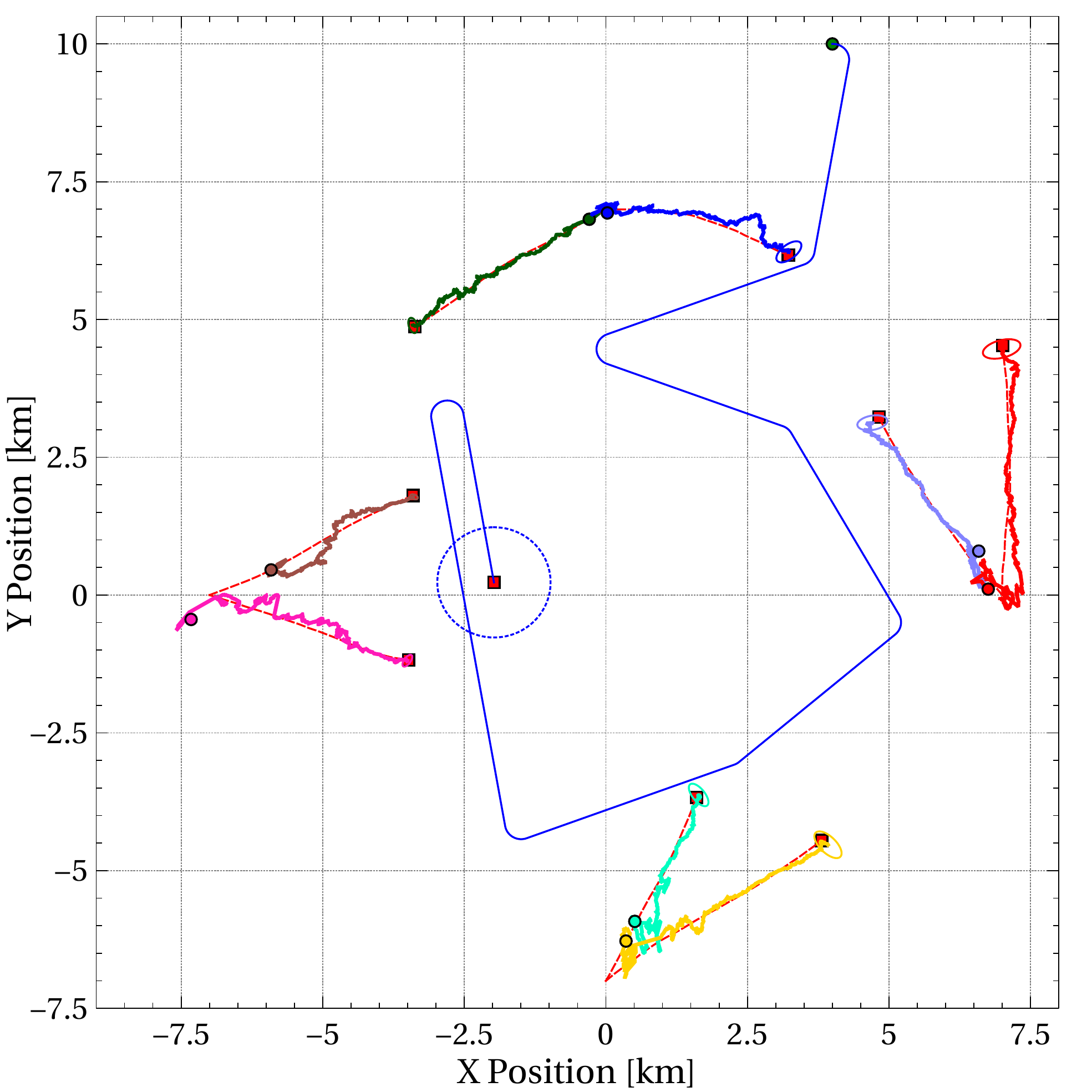}
\par\end{centering}

\caption{Scenario 2 - Typical sensor trajectory under the proposed control
scheme, along with the true and estimated target trajectories.}
\label{f:Scenario_2_Example}
\end{figure}

This scenario consists of 8 targets at various locations and moving
in different directions, and unlike the previous scenario, the best
path for the sensor to take is not immediately obvious. The scenario
geometry is shown in Figure \ref{f:Scenario_2_Example}, which also
depicts one of the typical sensor trajectories obtained during the
100 Monte Carlo runs. The starting location for the sensor is fixed
near the top right-hand corner for all runs, and in the particular
case shown in Figure \ref{f:Scenario_2_Example}, it moves around
the surveillance area, appearing to visit each pair of targets in
sequence.

\begin{figure}[H]
\begin{centering}
\includegraphics[width=1\columnwidth]{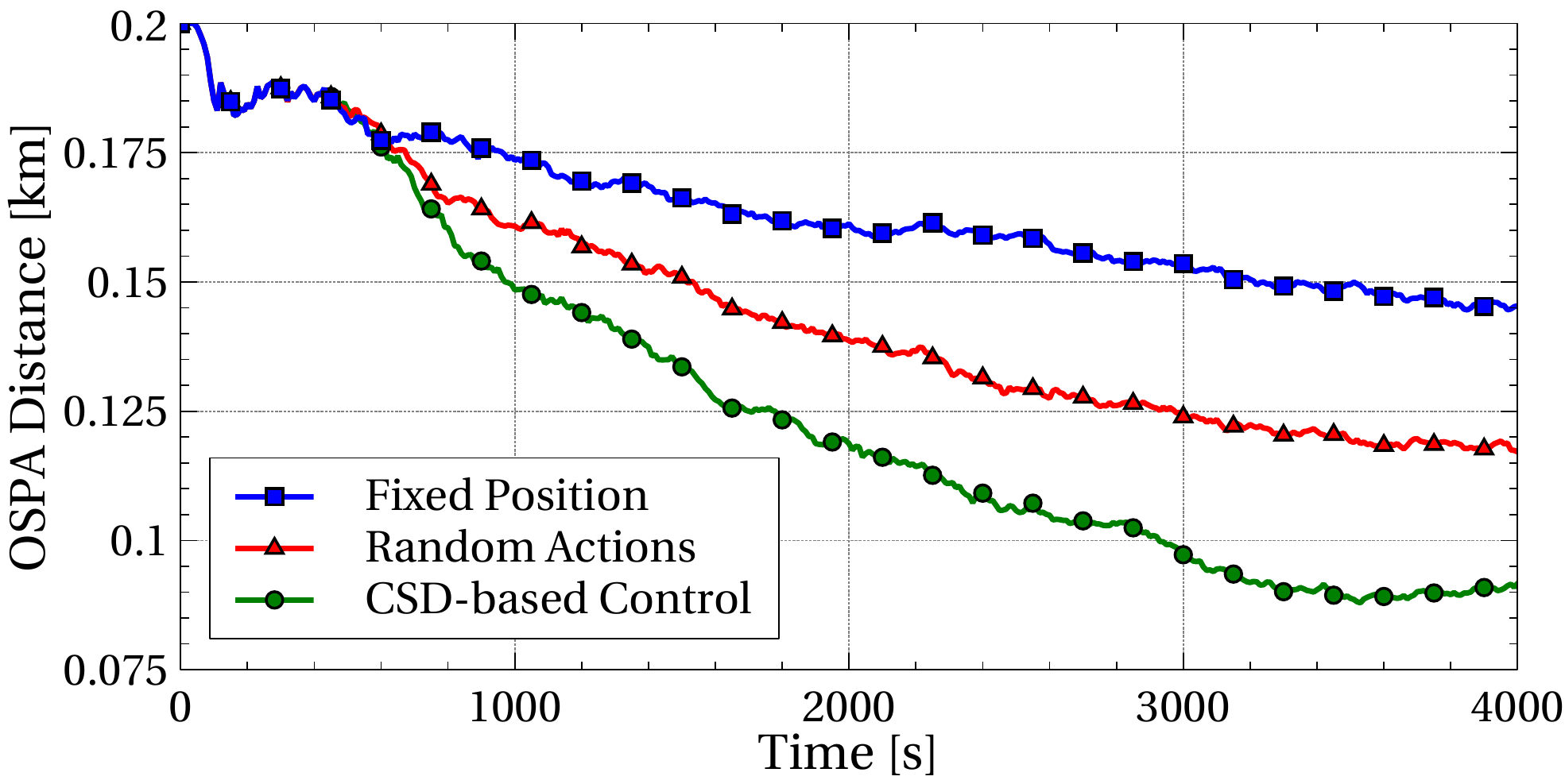}
\par\end{centering}

\caption{Scenario 2 - Comparison of tracking performance for fixed sensor location,
randomly chosen actions, and actions chosen using the Cauchy-Schwarz
divergence (CSD) based control scheme.}
\label{f:Scenario_2_OSPA}

\end{figure}

\begin{figure}[H]
\begin{centering}
\includegraphics[width=0.85\columnwidth]{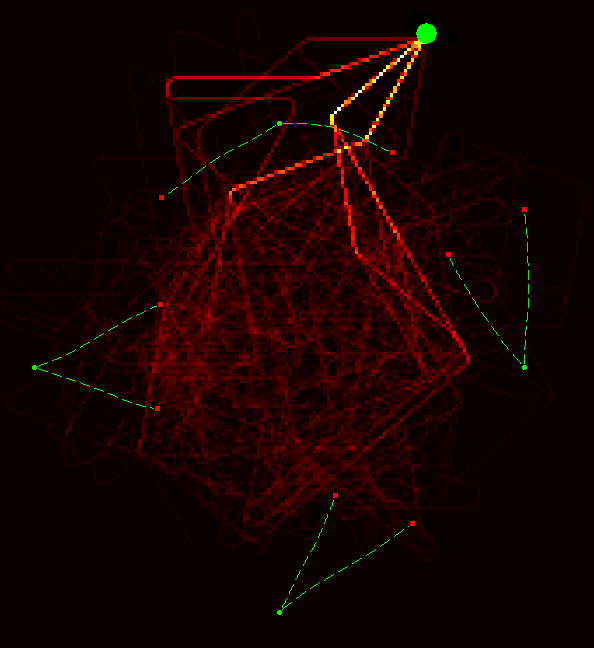}
\par\end{centering}

\caption{Scenario 2 - Heatmap showing the control behaviour over 100 Monte
Carlo runs.}

\label{f:Scenario_2_Heatmap}
\end{figure}

Figure \ref{f:Scenario_2_OSPA} shows a comparison of the OSPA distance
obtained for the cases of fixed sensor location, random actions, and
with the proposed control strategy. The sensor with fixed position
performs worst, because it has difficulty tracking the targets near
the bottom of the surveillance region due to their large distance
from the sensor. Moving with randomised actions improves the performance,
because despite the randomness of the chosen trajectories, the sensor
still has the opportunity to move closer to the far-away targets.
The proposed control strategy outperforms both the fixed and randomly
moving sensors, as indicated by the lower OSPA distance. Due to the
stochastic nature of the problem, the exact behaviour observed in
Figure \ref{f:Scenario_2_OSPA} is not necessarily replicated on every
run. However the sensor generally moves around the centre of the surveillance
region and attempts to visit each target in sequence. This can be
observed in Figure \ref{f:Scenario_2_Heatmap}, which shows that the
sensor spends most of the time moving around between the targets. 

\section{Conclusion\label{s:Conclusion}}

In this paper we have proposed two useful properties of generalized
labeled multi-Bernoulli models; an analytical form for the Cauchy-Schwarz
divergence between two GLMBs, and an analytical form for the void
probability functional of a GLMB. These properties have applications
in areas including GLMB mixture reduction, situational awareness,
and sensor management. Here we demonstrated their use in a sensor
management application, in which the goal was to plan a sensor trajectory
that optimizes the error performance in a multi-target tracking scenario.
The problem was formulated as a constrained POMDP, with a reward function
based on the expected Cauchy-Schwarz divergence, and a constraint
based on the void probability, to ensure adequate separation between
the sensor and targets. The results showed that this method was highly
effective at reducing the multi-target estimation error, compared
to cases where the sensor was stationary or undergoing random actions.
This demonstrates that both the proposed Cauchy-Schwarz divergence
and void probability functional are versatile tools in multi-object
information theory.

\end{document}